\documentclass[aps,graphicx,twocolumn]{revtex4}

\usepackage{graphicx}
\usepackage{epstopdf}
\usepackage{verbatim}
\usepackage{amsmath}
\usepackage[colorlinks,linkcolor=blue,citecolor=blue,urlcolor=blue]{hyperref}
\usepackage[caption=false,font=scriptsize,labelfont=sf,textfont=sf]{subfig}

\begin{document}
\title{High-efficient long-distance device-independent quantum secret sharing based on single-photon sources}

\author{Qi Zhang$^{1}$, Cheng Zhang$^{2}$, Wei Zhong$^{3}$, Ming-Ming Du$^{2}$, Lan Zhou$^{1}$\footnote{Email address: zhoul@njupt.edu.cn}, Yu-Bo Sheng$^{2}$}
\address{
$^1$College of Science, Nanjing University of Posts and Telecommunications, Nanjing, Jiangsu 210023, China\\
$^2$College of Electronic and Optical Engineering, and College of Flexible Electronics (Future Technology), Nanjing University of Posts and Telecommunications, Nanjing, Jiangsu 210023, China\\
$^3$Institute of Quantum Information and Technology, Nanjing University of Posts and Telecommunications, Nanjing, Jiangsu 210003, China\\
}
\date{\today}

\begin{abstract}
Device-independent quantum secret sharing (DI QSS) relaxes security assumptions of experimental devices to provide the highest security level for QSS. Previous DI QSS protocols require to generate multi-partite entangled states and then implement the long-distance entanglement distribution. The extremely low generation probability of current multi-partite entanglement source and photon transmission loss largely limit DI QSS's practical key generation rate and secure photon transmission distance. In the paper, we propose a high-efficient long-distance DI QSS protocol based on practical nearly on-demand single-photon sources (SPSs), which adopts the single photons and heralded architecture to construct long-distance multi-partite entanglement channels. Our SPS DI QSS protocol can automatically eliminate the negative influence from photon transmission loss and largely increase the secure photon transmission distance. Moreover, we adopt active improvement strategies such as random key generation basis, postselection, and advanced postselection to further increase the practical key generation rate and lower the experimental requirement. Comparing with previous DI QSS protocols based on entanglement sources, our SPS DI QSS protocol can increase the practical communication efficiency by six orders of magnitude, and increase the secure photon transmission distance by at least 151 times. This SPS DI QSS protocol is feasible with the current experimental technologies. Our protocol makes it possible to realize high-efficient long-distance DI network in the near future.
\end{abstract}
\maketitle

\section{Introduction}
Quantum secure communication ensures the unconditional security of the information transmission process based on the basic principles of quantum mechanics. As an important branch of quantum secure communication, quantum secret sharing (QSS) is a typical multipartite cryptographic primitive. QSS enables the dealer to divide each secret key into several parts (subkeys) and to distribute each subkey to a player. The transmitted keys can be recovered only by all the players working together \cite{QSS1}. QSS is an important part of the future quantum communication network. In 1999, Hillery \emph{et al}. proposed the original QSS protocol based on Greenberger-Horne-Zeilinger (GHZ) states \cite{QSS1}. Since then, QSS has been widely studied in theory and experiment \cite{QSS2,QSS3,QSS4,QSS5,QSS6,QSS7,QSS8,QSS9,QSS10,QSS11,MDIQSS1,MDIQSS2,DMDIQSS1,DPS1,RR1,QSS12,QSSe1,QSSe2,QSSe3,QSSe4,QSSe5,QSSe6,QSSe7,QSSe8}.

Similar to other branches of quantum secure communication, the practical imperfect experimental devices may lead to security vulnerabilities in QSS. Device-independent (DI) paradigm based on the violation of Bell-type inequalities \cite{Bell,CHSH} provides a promising method to eliminate the security vulnerability caused by practical imperfect devices. DI paradigm treats all experimental devices at each user's location as a black box, and eliminates all additional assumptions on the experimental devices. Only two basic assumptions need to be obeyed, say, quantum physics is correct and the physical location of each communication user is secure. As a result, DI paradigm can provide the highest security level for the quantum communication protocols \cite{DIQRG1,DIQRG2,DIQKD,DIQKD1,DIQKD2}. The research on DI-type protocols started with DI quantum key distribution (QKD) \cite{DIQKD,DIQKD1,DIQKD2}. In the last few years, DI QKD has made significant theoretical developments \cite{DIQKD3,DIQKD4,DIQKD5,DIQKD6,DIQKD7,DIQKD8,DIQKD9,DIQKD10,DIQKD11,DIQKD12,DIQKD13,DIQKD14,DIQKD15,DIQKD16,DIQKD17,Heralded1,Heralded3}. Recently, DI QSS and DI quantum secure direct communication protocols were proposed by \cite{DIQSS1,DIQSS2,DIQSS3,DIQSDC1,DIQSDC2,Heralded2}. As a trade-off for high-security, DI-type protocols have a high global detection efficiency threshold (more than 90\%) and low noise tolerance threshold \cite{DIQKD1,DIQKD2,DIQKD3,DIQSS2,DIQSS3,DIQSDC1}, which makes it difficult for researchers to implement them by using existing techniques, especially for optical implementations. Benefiting from the active improvement strategies \cite{DIQKD6,DIQKD7,DIQKD8,DIQKD9,DIQKD10,DIQKD11,DIQKD12}, DI QKD achieved experimental breakthroughs in 2022 \cite{DIQKDe1,DIQKDe2,DIQKDe3}. The research on DI QSS is still been in the beginning stage. In 2019, Roy \emph{et al}. proposed the first DI QSS protocol with arbitrary even dimensions and proved its correctness and completeness under the assumption of causal independence of measurement devices \cite{DIQSS1}. In 2024, the DI QSS protocol based on the violation of Svetlichny inequality \cite{Svetlichny} was proposed by \cite{DIQSS2}, and DI QSS's performance in practical communication environments has been characterized. Since multiple entangled photons need to be transmitted to multiple remote users in DI QSS, its equipment performance requirements are extremely high. A series of active improvement strategies have been introduced in DI QSS \cite{DIQSS2,DIQSS3} to decrease the global detection efficiency threshold from the initial 96.32\% to 93.41\%, and improve the communication distance from 0.26 km to 1.41 km.

Previous DI QSS protocols \cite{DIQSS2,DIQSS3} adopt the cascaded spontaneous parametric down-conversion (SPDC) sources to generate multi-photon entanglement \cite{GHZ1} and distribute entangled photons to multiple users. However, there are two shortcomings in this experimental setting. Firstly, the entanglement generation by periodically poled potassium titanyl phosphate (ppKTP) crystals is probabilistic ($10^{-5}-10^{-3}$) \cite{GHZ2}, resulting in a quite low probability to generate multi-photon GHZ states. This factor would largely reduce DI QSS's practical key generation rate. Secondly, entangled photon pairs are vulnerable to transmission loss during long-distance entanglement distribution, and the nonlocal correlation among photons is degraded seriously. It largely limits DI QSS's secure photon transmission distance. These two shortcomings seriously limit DI QSS's demonstration and application.

In order to overcome these negative effects and improve DI QSS's practical performances, in the paper, we propose a high-efficient DI QSS protocol based on the single-photon source (SPS) and heralded multipartite entanglement channel construction. Here, we call the previous DI QSS protocols \cite{DIQSS2,DIQSS3} based on cascaded SPDC source as SPDC DI QSS protocols, while our current DI QSS protocol as SPS DI QSS protocol. Comparing with previous SPDC DI QSS protocols \cite{DIQSS2,DIQSS3}, this SPS DI QSS protocol has some advantages. First, practical SPS \cite{source,source1} can generate a single photon in an almost on-demand and efficient way. Current SPSs can maintain the purity and indistinguishability of the generated photons with the probability of above 99\% \cite{source3,source4}. Recently, the telecommunication wavelength single-photon sources based on InAs/GaAs quantum dots with a count rate up to 10 MHz was reported by \cite{source2}. The adoption of SPSs can largely increase DI QSS's practical key generation rate. Second, the emitted single photons are used to heraldically construct the long-distance multipartite entanglement channels by the GHZ state measurement (GSM), where the negative effect from photon transmission loss on the channel construction can be automatically eliminated \cite{Heralded1}. Benefiting to above features, the SPS DI QSS protocol can increase the practical key generation rate by six orders of magnitude and extend the maximum secure communication distance by 183 times. Moreover, we also combine active improvement strategies to further improve DI QSS's performance. Our DI QSS protocol is feasible under current experimental conditions. It provides a possible method to realize high-efficient, long-distance DI QSS in the future.

The paper is organized as follows. In Sec.~\ref{Section2}, we explain the basic principle of the high-efficient SPS DI QSS protocol. In Sec.~\ref{Section3}, we characterize the performance of SPS DI QSS protocol in practical noise environments. In Sec.~\ref{Section4}, we introduce a series of active improvement strategies to further improve SPS DI QSS's performance. Finally, we make some discussion and conclusion in Sec.~\ref{Section5}.

\section{High-efficient three-partite SPS DI QSS protocol}\label{Section2}
\begin{figure*}
\includegraphics[scale=0.37]{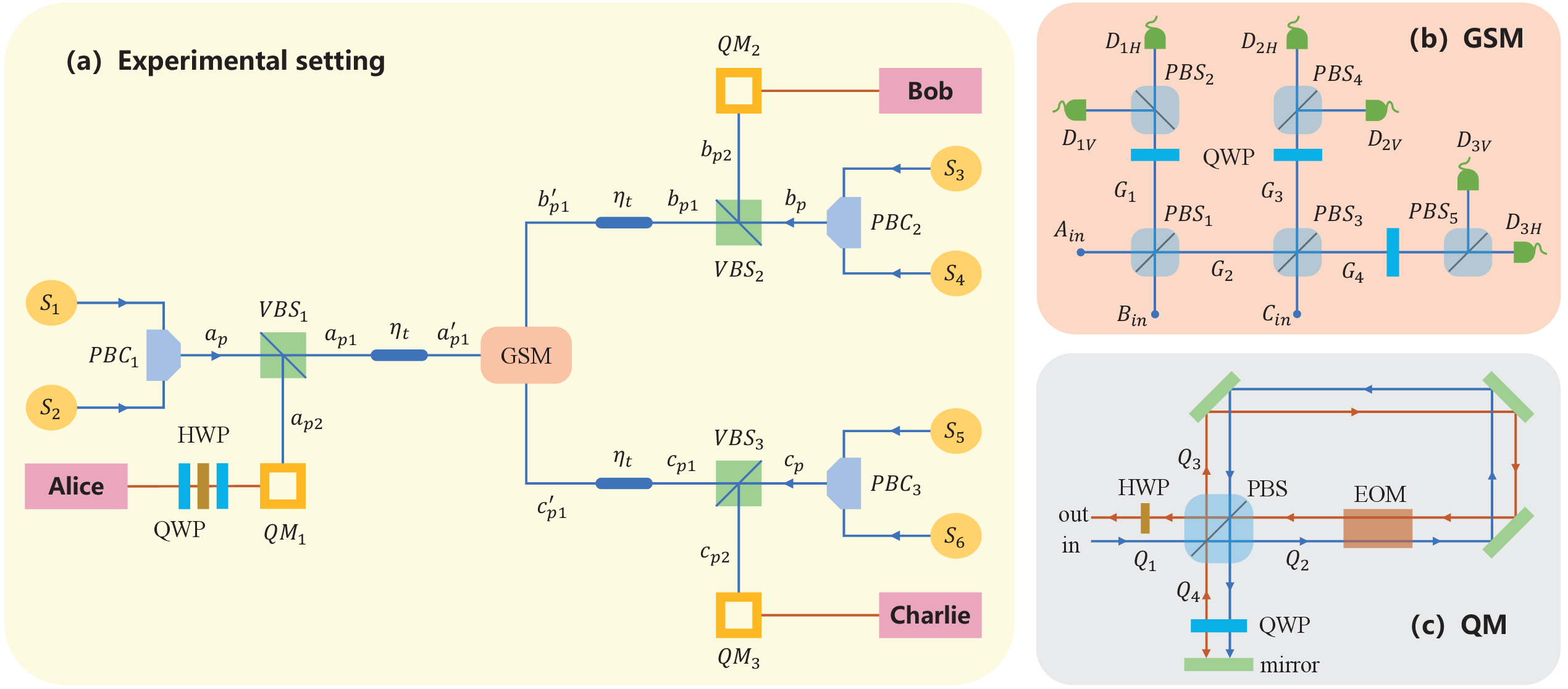}
\caption{(Color online) Schematic diagram of three-partite SPS DI QSS protocol. (a) Experimental setting of SPS DI QSS protocol. At each user's site, two photons with different polarization states generated from SPSs are coupled into the input port of the variable beam splitter (VBS) by the polarization beam combiner (PBC). Each photon may transmit the VBS with the probability of $T$ and pass through the lossy channel (transmission efficiency $\eta_t$) to reach the GHZ state measurement (GSM) module, or be reflected to the quantum memory (QM) module with the probability of $1-T$. HWP denotes the half-wave plate, which enables $|H\rangle\leftrightarrow|V\rangle$. QWP denotes the quarter-wave plate, which enables $|H\rangle\rightarrow\frac{1}{\sqrt{2}}(|H\rangle+|V\rangle)$ and $|V\rangle\rightarrow\frac{1}{\sqrt{2}}(|H\rangle-|V\rangle)$. (b) The structure of the GHZ state measurement module \cite{GSM}. PBS denotes the polarization beam splitter, which can totally transmit photons in $|H\rangle$ state and reflect photons in $|V\rangle$ state. $D_{1H}$, $D_{1V}$, $D_{2H}$, $D_{2V}$, $D_{3H}$, and $D_{3V}$ are single-photon detectors. (c) Quantum memory module \cite{QM1}. EOM is an electro-optical modulator which can control the storage and readout of photons by adjusting the polarization.}
\label{fig1}
\end{figure*}

The basic principle of SPS DI QSS protocol is shown in Fig.~\ref{fig1}. Before introducing SPS DI QSS, we first explain the heralded construction of the long-distance multipartite entanglement channels under linear optical conditions \cite{Heralded6} and the quantum memory (QM) \cite{QM1}.

\subsection{Heralded construction of the three-partite long-distance entanglement channels}\label{Section2.1}
As shown in Fig.~\ref{fig1} (a), Alice (Bob, Charlie) prepared two single photons in horizontal polarization states $|H\rangle$ and vertical polarization $|V\rangle$ from two SPSs $S_1$ and $S_2$ ($S_3$ and $S_4$, $S_5$ and $S_6$), respectively. The polarization beam combiners (PBCs) couples the two photons with different polarization state into the paths $a_p$, $b_p$, and $c_p$, respectively. Then the photons in paths $a_p$, $b_p$, and $c_p$ pass through the variable beam splitters (VBSs) with the transmittance of $T$. The quantum state of each two-photon system evolves to
\begin{eqnarray}\label{phiaphibphic}
|\phi_a\rangle&=&|H\rangle_{a_p}\otimes|V\rangle_{a_p}\nonumber\\
&\xrightarrow{VBS_1}&(\sqrt{T}|H\rangle_{a_{p1}}+\sqrt{1-T}|H\rangle_{a_{p2}})\nonumber\\
&&\otimes(\sqrt{T}|V\rangle_{a_{p1}}+\sqrt{1-T}|V\rangle_{a_{p2}}),\nonumber\\
|\phi_b\rangle&=&|H\rangle_{b_p}\otimes|V\rangle_{b_p}\nonumber\\
&\xrightarrow{VBS_2}&(\sqrt{T}|H\rangle_{b_{p1}}+\sqrt{1-T}|H\rangle_{b_{p2}})\nonumber\\
&&\otimes(\sqrt{T}|V\rangle_{b_{p1}}+\sqrt{1-T}|V\rangle_{b_{p2}}),\nonumber\\
|\phi_c\rangle&=&|H\rangle_{c_p}\otimes|V\rangle_{c_p}\nonumber\\
&\xrightarrow{VBS_3}&(\sqrt{T}|H\rangle_{c_{p1}}+\sqrt{1-T}|H\rangle_{c_{p2}})\nonumber\\
&&\otimes(\sqrt{T}|V\rangle_{c_{p1}}+\sqrt{1-T}|V\rangle_{a_{p2}}).
\end{eqnarray}

In order to successfully establish the long-distance entanglement channels, we only select the case where one photon transmits the VBS and the other photon is reflected at each user's site. The transmitted photons in paths $a_{p1}$, $b_{p1}$, and $c_{p1}$ pass through the lossy quantum channels with the transmission efficiency $\eta_t$ to the spatial modes $a'_{p1}$, $b'_{p1}$, and $c'_{p1}$, and the reflected photons in paths $a_{p2}$, $b_{p2}$, and $c_{p2}$ enter the QM modules. Here, we consider a symmetric case where the distance from the GSM module to each user is equal. Then, the whole quantum state $|\Phi\rangle=|\phi_a\rangle\otimes|\phi_b\rangle\otimes|\phi_c\rangle$ collapses as
\begin{eqnarray}\label{collapsePhi}
|\Phi\rangle&\rightarrow&\sqrt{\eta_tT(1-T)}\left(|H\rangle|V\rangle+|V\rangle|H\rangle\right)_{a_{p1}'a_{p2}}\nonumber\\
&\otimes&\sqrt{\eta_tT(1-T)}\left(|H\rangle|V\rangle+|V\rangle|H\rangle\right)_{b_{p1}'b_{p2}}\nonumber\\
&\otimes&\sqrt{\eta_tT(1-T)}\left(|H\rangle|V\rangle+|V\rangle|H\rangle\right)_{c_{p1}'c_{p2}}.
\end{eqnarray}

Then, the photons in spatial modes $a_{p1}'$, $b_{p1}'$, and $c_{p1}'$ enter the GSM module at the fourth party David's site from $A_{in}$, $B_{in}$ and $C_{in}$, respectively. The structure of the linear optical GSM module is shown in Fig.~\ref{fig1} (b) \cite{GSM}. The eight three-photon polarization GHZ states have the forms of
\begin{eqnarray}\label{GHZ}
|GHZ_1^\pm\rangle&=&\frac{1}{\sqrt{2}}\left(|HHH\rangle\pm |VVV\rangle\right),\nonumber\\
|GHZ_2^\pm\rangle&=&\frac{1}{\sqrt{2}}\left(|HHV\rangle\pm |VVH\rangle\right),\nonumber\\
|GHZ_3^\pm\rangle&=&\frac{1}{\sqrt{2}}\left(|HVH\rangle\pm |VHV\rangle\right),\nonumber\\
|GHZ_4^\pm\rangle&=&\frac{1}{\sqrt{2}}\left(|HVV\rangle\pm |VHH\rangle\right).
\end{eqnarray}

The photons successively pass through the polarization beam splitter (PBS) and quarter-wave plate (QWP), and will be detected by single-photon detector $D_{1H}$, $D_{1V}$, $D_{2H}$, $D_{2V}$, $D_{3H}$, and $D_{3V}$. The GSM with linear optical elements can only distinguish two of the eight GHZ states. The clicks from $D_{1H}D_{2H}D_{3H}$, $D_{1H}D_{2V}D_{3V}$, $D_{1V}D_{2H}D_{3V}$ or $D_{1V}D_{2V}D_{3H}$ herald $|GHZ_1^+\rangle$, while the clicks from $D_{1H}D_{2V}D_{3V}$, $D_{1H}D_{2H}D_{3V}$, $D_{1V}D_{2H}D_{3H}$ or $D_{1H}D_{2V}D_{3H}$ herald $|GHZ_1^-\rangle$.

After the GSM, the state in Eq.~\eqref{collapsePhi} becomes
\begin{eqnarray}\label{collapsePhi1}
|\Phi_1\rangle&=&\left(\sqrt{\eta_tT(1-T)}\right)^3\nonumber\\
&\times&(|GHZ_1^+\rangle_{A_{in}B_{in}C_{in}}|GHZ_1^+\rangle_{QM_1QM_2QM_3}\nonumber\\
&-&|GHZ_1^-\rangle_{A_{in}B_{in}C_{in}}|GHZ_1^-\rangle_{QM_1QM_2QM_3}\nonumber\\
&+&|GHZ_2^+\rangle_{A_{in}B_{in}C_{in}}|GHZ_2^+\rangle_{QM_1QM_2QM_3}\nonumber\\
&-&|GHZ_2^-\rangle_{A_{in}B_{in}C_{in}}|GHZ_2^-\rangle_{QM_1QM_2QM_3}\nonumber\\
&+&|GHZ_3^+\rangle_{A_{in}B_{in}C_{in}}|GHZ_3^+\rangle_{QM_1QM_2QM_3}\nonumber\\
&-&|GHZ_3^-\rangle_{A_{in}B_{in}C_{in}}|GHZ_3^-\rangle_{QM_1QM_2QM_3}\nonumber\\
&+&|GHZ_4^+\rangle_{A_{in}B_{in}C_{in}}|GHZ_4^+\rangle_{QM_1QM_2QM_3}\nonumber\\
&-&|GHZ_4^-\rangle_{A_{in}B_{in}C_{in}}|GHZ_4^-\rangle_{QM_1QM_2QM_3}).
\end{eqnarray}

From Eq.~\eqref{collapsePhi1}, it can be seen that the photons in QMs establish the long-distance GHZ entanglement $|GHZ_1^\pm\rangle_{QM_1QM_2QM_3}$ by the GSM. When the GSM result is $|GHZ_1^-\rangle$, Alice performs the phase flipping operation to transform $|GHZ_1^-\rangle_{QM_1QM_2QM_3}$ to $|GHZ_1^+\rangle_{QM_1QM_2QM_3}$. As a result, when the GSM is successful and all QMs are loaded by photons, Alice, Bob and Charlie can deterministically construct the three-partite long-distance GHZ entanglement channels. The probability $P_d$ of successfully establishing $|GHZ_1^+\rangle_{QM_1QM_2QM_3}$ is given by
\begin{eqnarray}\label{pd}
P_d=2\eta_t^3T^3(1-T)^3.
\end{eqnarray}
To maximize the probability of successfully constructing the GHZ state entanglement channel, the transmittance of the VBS is set as $T=0.5$. In this case, we can obtain $P_d=\eta_t^3/32$.

It is noticed that when the GSM fails or any QM not loaded by photon, Alice, Bob, and Charlie have to discard the photons in QMs. As a result, the effect from photon transmission loss case can be automatically eliminated.
\subsection{Quantum memory module}\label{Section2.2}
Our DI QSS protocol adopts the feasible all-optical, polarization-insensitive storage loop \cite{QM1} as the QM. The structure of the QM is shown in Fig.~\ref{fig1} (c).

We take the storage and readout processes of Alice's photon in QM as an example. The photon in spatial modes $a_{p2}$ enters the $QM_{1}$ at the \emph{in} port and exits at the \emph{out} port. Here, the electro-optical modulator (EOM) in the ON state can flip the photon's polarization state, while the EOM in the OFF state does not change the photon's polarization state. In this way, Alice can realize the photon storage and readout from the storage loop by controlling the ON-OFF of the EOM. When a photon enters the QM, the EOM state is switched from OFF to ON. If a photon should be read out, the EOM state is switched from ON to OFF. When the photon exits the storage loop from the spatial mode $Q_4$, the combination of QWP and mirror causes the photon to pass through the QWP twice, which rotates the photon polarization state by 90 degrees, \emph{i.e.}, $|H\rangle(|V\rangle)\xrightarrow{double~QWP}|V\rangle(|H\rangle)$. Then, the photon will reenter the storage loop.

The principle details of the QM are shown as follows. Case 1: When the input photon state is in $|H\rangle$, the storage and readout processes can be written as
\begin{eqnarray}\label{QMcase1}
&&|H\rangle_{Q_1}^{in}\xrightarrow{PBS}|H\rangle_{Q_2}\xrightarrow{EOM(OFF)}|H\rangle_{Q_3}\xrightarrow{PBS}|H\rangle_{Q_4}\nonumber\\
&&\xrightarrow{double~QWP}|V\rangle_{Q_4}\xrightarrow{PBS}|V\rangle_{Q_2}\xrightarrow{EOM(ON)}|H\rangle_{Q_3}\nonumber\\
&&\xrightarrow{storing~photon}\cdots\xrightarrow{double~QWP}|V\rangle_{Q_4}\xrightarrow{PBS}|V\rangle_{Q_2}\nonumber\\
&&\xrightarrow{EOM(OFF)}\cdots\xrightarrow{read~out~photon}|V\rangle_{Q_3}\xrightarrow{PBS}|V\rangle_{Q_1}\nonumber\\
&&\xrightarrow{HWP}|H\rangle_{Q_1}^{out}.
\end{eqnarray}

Case 2: When the input photon state is in $|V\rangle$, the storage and readout processes are
\begin{eqnarray}\label{QMcase2}
&&|V\rangle_{Q_1}^{in}\xrightarrow{PBS}|V\rangle_{Q_3}\xrightarrow{EOM(OFF)}|V\rangle_{Q_2}\xrightarrow{PBS}|V\rangle_{Q_4}\nonumber\\
&&\xrightarrow{double~QWP}|H\rangle_{Q_4}\xrightarrow{PBS}|H\rangle_{Q_3}\xrightarrow{EOM(ON)}|V\rangle_{Q_2}\nonumber\\
&&\xrightarrow{storing~photon}\cdots\xrightarrow{double~QWP}|H\rangle_{Q_4}\xrightarrow{PBS}|H\rangle_{Q_3}\nonumber\\
&&\xrightarrow{EOM(OFF)}\cdots\xrightarrow{read~out~photon}|H\rangle_{Q_2}\xrightarrow{PBS}|H\rangle_{Q_1}\nonumber\\
&&\xrightarrow{HWP}|V\rangle_{Q_1}^{out}.
\end{eqnarray}

It can be found that when the photon is read out from the QM, its polarization is flipped. In this way, we should use an HWP to recover its polarization feature.

\subsection{The SPS DI QSS protocol}\label{Section2.3}
Similar to all DI-type protocols, our DI QSS protocol only needs to obey two basic assumptions. Firstly, quantum physics is correct, and the eavesdropper (Eve) obeys the basic laws of quantum mechanics. Secondly, the physical locations of Alice, Bob, and Charlie are secure. Meanwhile, the three users must be legitimate and honest during the key generation process. The main steps of SPS DI QSS protocol are as follows:

\textbf{Step 1 Construction of long-distance entanglement channels.} Alice, Bob, and Charlie prepare $N$ ($N\to \infty$) single-photon pairs in $|H\rangle\otimes|V\rangle$ using the SPSs, respectively. Then, they construct $N_1$ pairs of three-partite long-distance GHZ states in $|GHZ_1^{+}\rangle$ as shown in Sec.~\ref{Section2.1}, where $N_1=NP_d$. The entangled photons stored in the quantum memories $QM_1$, $QM_2$ and $QM_3$ form the photon sequences $L_{A1}$, $L_{B1}$ and $L_{C1}$, respectively.

\textbf{Step 2 Photon measurements.} Alice, Bob, and Charlie independently and randomly choose measurement bases to measure the received photons. Where Alice has two measurement bases, $A_1=\sigma_x$ and $A_2=\sigma_y$. Bob has three measurement bases, including $B_1=\sigma_x$, $B_2=\frac{\sigma_x-\sigma_y}{\sqrt{2}}$, and $B_2=\frac{\sigma_x+\sigma_y}{\sqrt{2}}$. Charlie has two measurement bases, $C_1=\sigma_x$ and $C_2=-\sigma_y$. Measurement results of three users are denoted as $a_i$, $b_j$ and $c_k$ ($i,k\in\{1,2\}, j\in\{1,2,3\}$), where $a_i,b_j,c_k\in\{-1,+1\}$. The probability that a photon suffers the local loss during the measurement process is $\bar{\eta}_l=1-\eta_l$, and we labeled the detector no-click event as $\perp$.

\textbf{Step 3 Security checking and key generation.} After all photons are measured, Alice, Bob, and Charlie announce the measurement bases for each photon in the $L_{A1}$, $L_{B1}$, and $L_{C1}$ sequences in turn. Based on the choice of measurement bases, there are three scenarios.

Case 1: Security analysis is performed when Bob chooses $B_2$ or $B_3$. In this case, all users publish their measurement results for estimating the Svetlichny polynomial $S_{ABC}$ \cite{Svetlichny}, as well as the CHSH polynomial $S$ between Alice's and Bob's results. $S_{ABC}$ can be calculated as
\begin{eqnarray}\label{Svetlichnypolynomial}
S_{ABC}&=\langle a_{1}b_{2}c_{2}\rangle+\langle a_{1}b_{3}c_{1}\rangle+\langle a_{2}b_{2}c_{1}\rangle-\langle a_{2}b_{3}c_{2}\rangle \nonumber\\
&+\langle a_{2}b_{3}c_{1}\rangle+\langle a_{2}b_{2}c_{2}\rangle+\langle a_{1}b_{3}c_{2}\rangle-\langle a_{1}b_{2}c_{1}\rangle.
\end{eqnarray}
Here, $\langle a_{i}b_{j}c_{k}\rangle$ is the expected value of the tripartite measurement results with the form of  $\langle a_{i}b_{j}c_{k}\rangle =P(a_{i}b_{j}c_{k}=1)-P(a_{i}b_{j}c_{k}=-1)$.

For the three-photon GHZ state, Svetlichny polynomial $S_{ABC}$ can be simplified by the CHSH polynomials as \cite{nonlocal1}
\begin{eqnarray}\label{Svetlichnypolynomial2}
S_{ABC}=\langle S_{AB}c_{2}\rangle+\langle S_{AB}'c_{1}\rangle.
\end{eqnarray}
Here, $S_{AB}$ and $S_{AB}'$ are the general CHSH polynomial and its equivalent variation between Alice's and Bob's measurement results with the form of
\begin{eqnarray}\label{CHSH}
S_{AB}&=&\langle a_{1}b_{2}\rangle+\langle a_{2}b_{2}\rangle+\langle a_{1}b_{3}\rangle-\langle a_{2}b_{3}\rangle,\nonumber\\
S_{AB}'&=& \langle a_{2}b_{3}\rangle+\langle a_{2}b_{2}\rangle+\langle a_{1}b_{3}\rangle-\langle a_{1}b_{2}\rangle,
\end{eqnarray}
with $\langle a_{i}b_{j}\rangle=P(a_{i}b_{j}=1)-P(a_{i}b_{j}=-1)$.

For simplicity, we define $S$ to represent $S_{AB}$ or $S_{AB}'$ based on the following rule
 \begin{equation}\label{Bell}
	S=\begin{cases}
	   S_{AB}' ,& \text{if     $c_1=+1$},\\
	   -S_{AB}' ,& \text{if     $c_1=-1$},\\
        S_{AB} ,& \text{if     $c_2=+1$},\\
	   -S_{AB} ,& \text{if     $c_2=-1$}.
	\end{cases}
\end{equation}

The violation of Svetlichny inequality ($S_{ABC}>4$) implies the existence of the genuine tripartite nonlocality. In this way, when $2<S\leq 2\sqrt{2}$ (equivalent to $4<S_{ABC}\leq 4\sqrt{2}$), it implies that three users' corresponding measurement results are nonlocal correlation \cite{nonlocal1}. In this scenario, the users can estimate the key leakage rate based on $S$ ($S_{ABC}$) value, so that they ensure that the photon transmission process is secure and the communication continues. On the contrary, when $S\leq 2$ (equivalent to $S_{ABC}\leq 4$), it implies that three users' measurement results only have classical correlation. In this scenario, the users cannot estimate the key leakage rate, so that the photon transmission process is unsecure, and the communication has to be terminated.

Case 2: When three users choose the measurement basis combination is $\{A_1B_1C_1\}$ or $\{A_2B_1C_2\}$, they each preserve the corresponding measurement result as the key bit. The measurement result $+1$ is labeled as key bit $0$, and $-1$ is labeled as key bit $1$. Similar as Refs. \cite{QSS1,DIQSS2}, the encoding rule is $k_A=k_B\oplus k_C$, where $k_A$, $k_B$, and $k_C$ denote key bits of Alice, Bob, and Charlie, respectively. Alice randomly announces some key bits. Bob and Charlie announce their corresponding key bits to estimate the qubit error rate (QBER) $\delta$. The remaining unannounced key bits form their raw key.

Case 3: For measurement basis combination is $\{A_1B_1C_2\}$ or $\{A_2B_1C_1\}$, three users have to discard the corresponding measurement results.

\textbf{Step 4 Error correction and private amplification.} Alice, Bob, and Charlie repeat the above steps until they obtain sufficient raw keys. Then, they perform classical error correction and private amplification on the obtained raw keys to finally form a series of secure keys.

\textbf{Step 5 Reconstruction of the secret keys.} According to the encoding rules, Charlie publishes his subkey $K_C$, and Bob combines his subkey $K_B$ to reconstruct the Alice's key $K_A$.

\section{Performance characterization in practical noise environments}\label{Section3}
The DI-type protocol does not limit the ability of Eve, assuming only that he obeys the fundamental principle of quantum mechanics, and he can even control single-photon sources and the user's measurement devices. In the practical implementation, we cannot make any security assumptions about all devices. The users only rely on the tripartite nonlocality to prove that the output results are truly randomized for Eve, thus guaranteeing the security of the transmitted key. All users statistically observe the correlation between the measurement basis choices and results by estimating Svetlichny polynomial (CHSH polynomials), thereby bounding Eve's knowledge about the key. In Step 2, the users control the probability of choosing the measurement basis. For example, the users control half of the entanglement photon pairs to be used for security checking.  We consider the defense of DI QSS against collective attacks, \emph{i.e.}, Eve takes the same attack on each subsystem, so that the knowledge of Eve can be reduced to a single round for analysis.

In this section, we estimate the performance of our DI QSS protocol with SPSs in the practical noise environment. Our SPS DI QSS protocol uses the measurement results of two basis combinations ($\{A_1B_1C_1\}$ and $\{A_2B_1C_2\}$) to generate the key bits. We assume that Alice chooses $A_1$ and $A_2$ measurement basis with probabilities $p$ and $\bar{p}=1-p$, and Charlie chooses $C_1$ and $C_2$ with $p$ and $\bar{p}$, respectively. In the practical implementation, the value of $p$ and $\bar{p}$ are set as $p=\bar{p}=50\%$. In the asymptotic limit of a large number of rounds, the total secure key rate $R_\infty$ which is expressed as the ratio of the extractable key length to the number of measurement rounds, is given by
\begin{eqnarray}\label{rinfty1}
R_\infty=p^2 r_{111}+\bar{p}^2 r_{212}=\left(p^2+\bar{p}^2\right)\left(\lambda r_{111}+\bar{\lambda}r_{212}\right),
\end{eqnarray}
where $\lambda=p^2/(p^2+\bar{p}^2)$ represents the matching weight of the first key generating basis combination $\{A_1B_1C_1\}$, and $\bar{\lambda}=1-\lambda$ is the matching weight of the second key generating basis combination $\{A_2B_1C_2\}$. The corresponding secure key rates for the two key generating basis combinations $\{A_1B_1C_1\}$ and $\{A_2B_1C_2\}$ are given by the Devetak-Winter rate \cite{DWrate1,DWrate2}, with the form of
\begin{eqnarray}\label{r111212}
r_{111}&=& H\left(A_1|E\right)-H\left(A_1|B_1,C_1\right),\nonumber\\
r_{212}&=& H\left(A_2|E\right)-H\left(A_2|B_1,C_2\right),
\end{eqnarray}
where $H(|)$ is the von Neumann conditional entropy. $H(A_1|E)$ and $H(A_2|E)$ quantify Eve's uncertainty about Alice's key corresponding to the basis combinations $\{A_1B_1C_1\}$ and $\{A_2B_1C_2\}$, respectively. $H(A_1|B_1,C_1)$ and $H(A_2|B_1,C_2)$ quantify the key error rate of Bob and Charlie corresponding to the basis combinations $\{A_1B_1C_1\}$ and $\{A_2B_1C_2\}$, respectively.

Therefore, Eq.~\eqref{rinfty1} is rewritten as
\begin{eqnarray}\label{rinfty2}
R_\infty&=&\left(p^2+\bar{p}^2\right)\Big[H\left(A|E\right)\nonumber\\
&-&\left(\lambda H\left(A_1|B_1,C_1\right)+\bar{\lambda}H\left(A_2|B_1,C_2\right)\right)\Big],
\end{eqnarray}
where $H(A|E)=\lambda H(A_1|E)+\bar{\lambda}H(A_2|E)$ is defined as the total key secrecy rate of Eve. In Ref.~\cite{DIQSS3}, we have analytically or numerically analyzed the lower bound on the total key secrecy rate with different probabilities $p$ is given by
\begin{eqnarray}\label{keysecrecyrate}
H\left(A|E\right)\ge g\left(\tilde{E}_\lambda(S)\right).
\end{eqnarray}
Where we define the function $g(x)=1-h(\frac{1}{2}+\frac{1}{2}x)$ and the binary Shannon entropy $h(x)=-x\log_{2}{x}-(1-x)\log_{2}{(1-x)}$. $\tilde{E}_\lambda(S)\equiv \sqrt{\tilde{E}_\lambda(S)^2}$, and $\tilde{E}_\lambda(S)^2$ is the maximum value obtained by numerically or analytically solving the optimization problem:
\begin{eqnarray}\label{maxproblem}
E_\lambda(S)^2=\max& s^2g^2+c^2h^2+2(2\lambda-1)scgh\varDelta, \nonumber\\
s.t.& cg+sh\geq S/2, \nonumber\\
& g^2 \leq 1, \nonumber\\
& h^2 \leq 1, \nonumber\\
& (1-g^2)(1-h^2)\geq g^2h^2\varDelta^2, \nonumber\\
& c^2+s^2=1, \nonumber\\
& \varDelta ^{2} \leq 1,
\end{eqnarray}
The specific calculations is shown in Ref.~\cite{DIQSS3}.

Although the heralded structure as described in Sec. \ref{Section2.1} eliminates the effects from photon transmission loss, photon local loss, and decoherence may still occur in storage and measurement by noise. These negative effects can severely break the entanglement and weaken the nonlocal correlation among users' measurement results. Here, we consider a simple white noise model, where the target GHZ state may degenerate into eight possible GHZ states \eqref{GHZ} with equal probability. Assume that each user successfully detects photon with the local efficiency $\eta_l$. In this way, the users share mixed states with the form of
\begin{eqnarray}\label{rhoABC1}
\rho_{ABC}&=&\eta^3_l\left(F|GHZ^+_1\rangle\langle GHZ^+_1|+\frac{1-F}{8}I\right)_{ABC}\nonumber\\
&+&\frac{1}{2}\eta^2_l\bar{\eta_l}\left(|HH\rangle\langle HH|+|VV\rangle\langle VV|\right)_{BC}\nonumber\\
&+&\frac{1}{2}\eta^2_l\bar{\eta_l}\left(|HH\rangle\langle HH|+|VV\rangle\langle VV|\right)_{AC}\nonumber\\
&+&\frac{1}{2}\eta^2_l\bar{\eta_l}\left(|HH\rangle\langle HH|+|VV\rangle\langle VV|\right)_{AB}\nonumber\\
&+&\frac{1}{2}\eta_l\bar{\eta_l}^2\left(|H\rangle\langle H|+|V\rangle\langle V|\right)_{A}\nonumber\\
&+&\frac{1}{2}\eta_l\bar{\eta_l}^2\left(|H\rangle\langle H|+|V\rangle\langle V|\right)_{B}\nonumber\\
&+&\frac{1}{2}\eta_l\bar{\eta_l}^2\left(|H\rangle\langle H|+|V\rangle\langle V|\right)_{C}\nonumber\\
&+&\bar{\eta_l}^3|vac\rangle\langle vac|.
\end{eqnarray}
Here, the detector no-click probability $\bar{\eta_l}=1-\eta_l$, $F$ is the fidelity of the target GHZ state $|GHZ^+_1\rangle$, and the unitary matrix $I$ is composed by the density matrices of the eight possible noisy GHZ states, and $|vac\rangle$ is the vacuum state.

We define the QBERs caused by the decoherence corresponding to the basis combinations $\{A_1B_1C_1\}$ and $\{A_2B_1C_2\}$ are $Q_{111}$ and $Q_{212}$, respectively. According to the coding rules \cite{DIQSS3}, $Q_{111}$ and $Q_{212}$ can be given by
\begin{eqnarray}\label{QBERQ1}
Q_{111}=Q_{212}=\frac{1}{2}\left(1-F\right)\eta_l^3.
\end{eqnarray}
As the basis combinations $\{A_1B_1C_1\}$ and $\{A_2B_1C_2\}$ occur with the same probability, the total QBER ($Q_{1}$) caused by the decoherence is $Q_{1}=Q_{111}=Q_{212}$.

To close the detection efficiency loophole of the Bell test, each user defines an additional output $\perp$ for no-click events. They define that an error occurs when there is one or more $\perp$ output results. Therefore, the QBER ($Q_2$) caused by photon local loss is given by
\begin{eqnarray}\label{QBERQ2}
Q_2&=&1-\eta^3_l.
\end{eqnarray}

With the noise model in Eq.~\eqref{rhoABC1}, the total QBER ($\delta$) can be calculated as
\begin{eqnarray}\label{lossdelta1}
\delta&=&Q_{1}+Q_{2}=\frac{1-F}{2}\eta_l^3+1-\eta_l^{3}\nonumber\\
&=&1-\frac{1}{2}\eta_l^3-\frac{1}{2}\eta_l^3F.
\end{eqnarray}

In this way, in above noise model, we can obtain $H\left(A_1|B_1,C_1\right)$ and $H\left(A_2|B_1,C_2\right)$ as
\begin{eqnarray}\label{HA1B1C1}
H\left(A_1|B_1,C_1\right)=h(Q_{111})=h(\delta),\nonumber\\
H\left(A_2|B_1,C_2\right)=h(Q_{212})=h(\delta).
\end{eqnarray}

Combined Eq.~\eqref{HA1B1C1} with Eq.~\eqref{keysecrecyrate}, the lower bound of the asymptotic secure key rate $R_\infty$ of our SPS DI QSS protocol in Eq.~\eqref{rinfty2} can be rewritten as
\begin{eqnarray}\label{keyrate}
R_\infty\ge\left(p^2+\bar{p}^2\right)\left[g\left(\tilde{E} _\lambda(S)\right)-h(\delta)\right].
\end{eqnarray}

Similar to the original DI QSS protocol \cite{DIQSS2}, the theoretical value of the CHSH polynomial $S$ between Alice's and Bob's measurement is
\begin{eqnarray}\label{noiseS}
S=2\sqrt{2}F\eta_l^3=2\sqrt{2}(2-\eta_l^3-2\delta).
\end{eqnarray}
When $2<S\leq 2\sqrt{2}$ (equivalently $4<S_{ABC}\leq 4\sqrt{2}$), three users ensure that there is nonlocal correlation between their measurement results.

We set the repetition frequency of the on-demand single photon source as $R_{rep}=10$ MHz \cite{source2}. The lossy channel's transmission efficiency is $\eta_t=10^{-\frac{\alpha d}{10}}$, where $d$ denotes the photon transmission distance, and $\alpha=0.2$ dB/km for a standard optical fiber. We define the local efficiency $\eta_l$ as the product of the efficiency $\eta_c$ of the photon-fiber coupling, the efficiency $\eta_m$ of the quantum memory, and the detection efficiency $\eta_d$ of the photon detector ($\eta_l=\eta_c\eta_m\eta_d$). In this way, we can obtain the practical key generation efficiency $E_c$ by
\begin{eqnarray}\label{EC}
E_c&=&(1-P_c)R_{rep}P_dR_\infty\geq \frac{\eta_t^3R_{rep}}{64}(p^2+\bar{p}^2)\nonumber\\
&&\left[g\left(\tilde{E} _\lambda(2\sqrt{2}F\eta_l^3)\right)-h(1-\frac{1}{2}\eta_l^3-\frac{1}{2}\eta_l^3F)\right].
\end{eqnarray}

\begin{figure}
\includegraphics[scale=0.34]{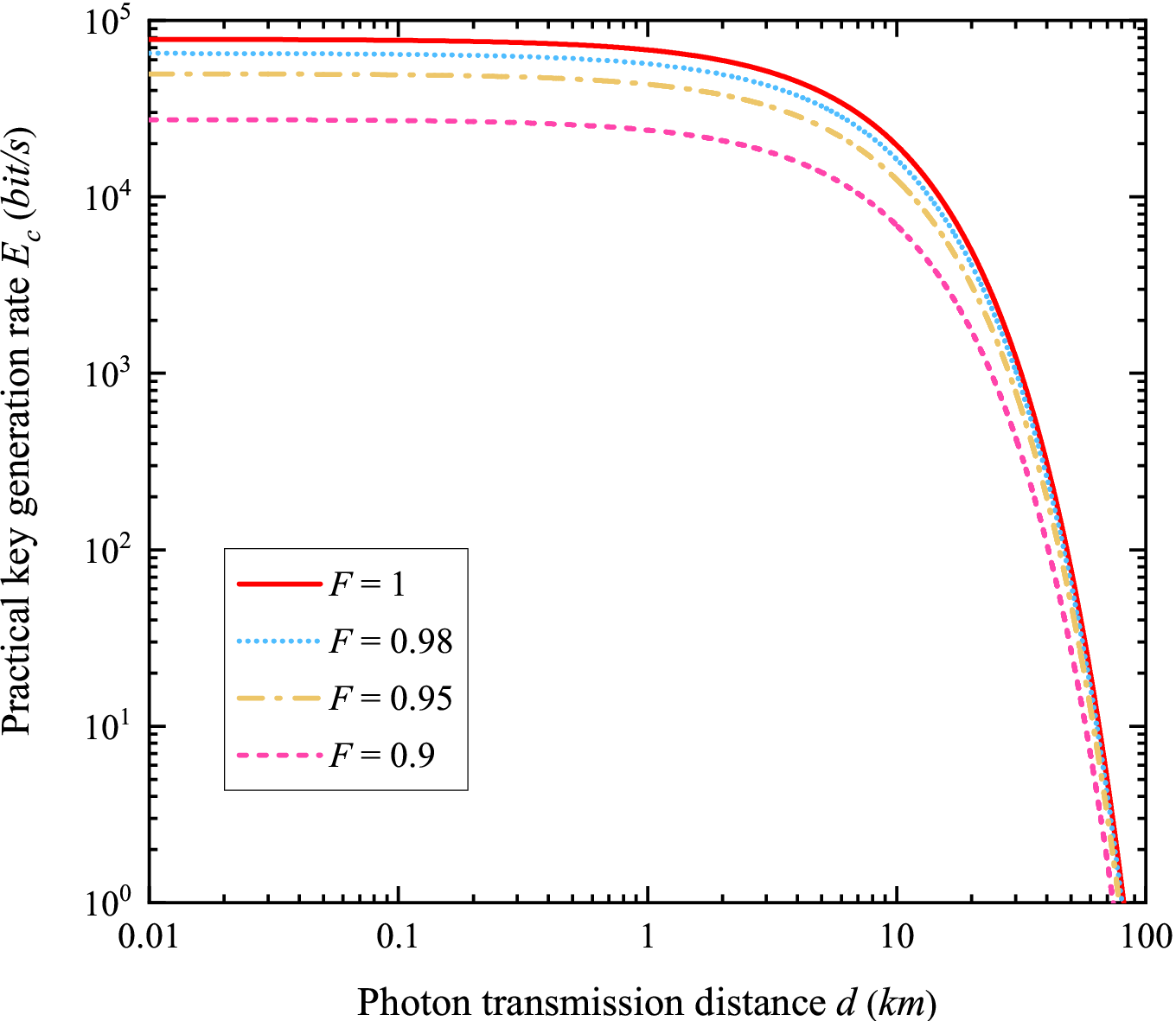}
\caption{(Color online) The practical key generation rate $E_c$ versus the photon transmission distance $d$ and the fidelity $F$, where Alice (Charlie) equally chooses $A_1$ and $A_2$ ($C_1$ and $C_2$) ($p=50\%$). We assume that the fidelity $F=1,0.98,0.95,0.9$, and the local efficiency $\eta_l=100\%$.}
\label{fig2}
\end{figure}

In Fig.~\ref{fig2}, we provide the lower bound of the practical key generation rate $E_c$ of the SPS DI QSS protocol versus the photon transmission distance $d$ and the fidelity $F$ when Alice (Charlie) equally chooses $A_1$ and $A_2$ ($C_1$ and $C_2$) ($p=50\%$) and the local efficiency $\eta_l=100\%$. As our protocol can eliminate the negative influence from photon transmission loss, the secure key rate $R_\infty$ independent of the photon transmission distance $d$. Only the probability $P_d$ of successfully constructing the multipartite entanglement channels is proportional to $\eta_{t}^3$. As a result, under the four cases in Fig.~\ref{fig2}, $E_c$ reduces with the growth of $\eta_{t}$, but is always positive. On the other hand, it seems that $E_{c}$ reduces with the decrease of the fidelity $F$. For obtaining the positive key generation rate, the threshold of $F$ is about 81.54\%. At the practical key generation rate of 1 bit/s, the secure photon transmission distance decreases from 81.55 km to 73.97 km when the fidelity decreases from $F=1$ to $F=0.9$.

\begin{figure}
\includegraphics[scale=0.34]{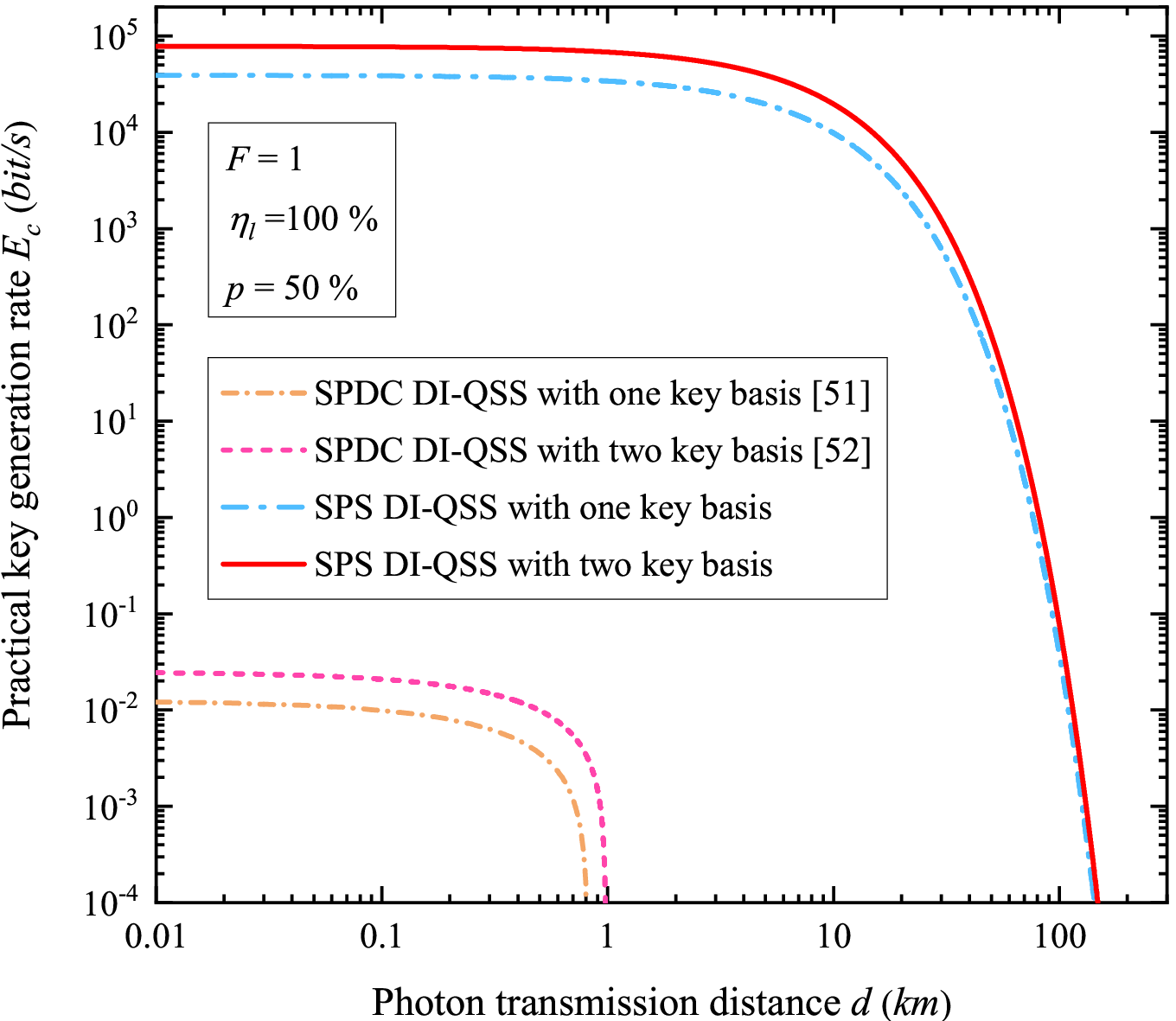}
\caption{(Color online) The comparison of the practical key generation rates of SPS DI QSS protocol and SPDC DI QSS protocols \cite{DIQSS2,DIQSS3}, with the fidelity $F=1$ and the local efficiency $\eta_l=100\%$. Here, we assume that the repetition frequencies of the SPDC source and SPS are both 10 MHz, Alice (Charlie) equally chooses $A_1$ and $A_2$ ($C_1$ and $C_2$) ($p=50\%$).}
\label{fig3}
\end{figure}

In Fig.~\ref{fig3}, we compare $E_c$ of the SPS DI QSS with the previous SPDC DI QSS protocols \cite{DIQSS2,DIQSS3} with $F=1$ and $\eta_l=100\%$. It is obvious that the practical key generation rate of the SPS DI QSS protocol is about six orders of magnitude of those in Ref.~\cite{DIQSS2,DIQSS3}. There are two main reasons. On one hand, previous SPDC DI QSS protocols \cite{DIQSS2,DIQSS3} adopt the central cascaded SPDC sources to directly generate three-photon GHZ states \cite{GHZ1} and distribute the entangled photons to multiple users. Suppose that each SPDC source generates the two-photon entangled state with the probability of $10^(-4)$, so that the GHZ state can be generated with the low probability of $10^{-8}$. Combing with the photon distribution, the total success probability of constructing the three-partite entanglement channels is $P_d^{SPDC}=10^{-8}\eta_t^3$. However, the SPSs can emit single photon nearly on-demand (we suppose the success probability of single photon emission is 100\%), based on the calculations in Sec.~\ref{Section2.1}, the success probability of constructing the three-partite entanglement channels reaches $P_d=\eta_t^3/32$, which is about six orders of magnitude higher than that of the SPDC DI QSS protocols. On the other hand, in SPDC DI QSS protocols \cite{DIQSS2,DIQSS3}, $S$ ($S_{ABC}$) value and the total QBER ($\delta$) are vulnerable to photon transmission loss, and thus their maximal secure photon transmission distance is quite low. In the SPS DI QSS protocol, benefiting from the heralded function of the GSM and QM, the negative influence of $\eta_{t}$ on $S$ ($S_{ABC}$) and $\delta$ can be automatically eliminated, which can largely extend the secure photon transmission distance and further increase the practical key generation rate. The SPDC DI QSS has quite limited maximal secure photon transmission distance \cite{DIQSS2,DIQSS3}, but the SPS DI QSS protocol's secure photon transmission distance can be infinite in theory. Here, we set $E_{c}=10^{-4}$, its secure photon transmission distance is about 148.21 km (about 183 times of 0.81 km in Ref.~\cite{DIQSS2} and 151 times of 0.98 km in Ref.~\cite{DIQSS3}). Meanwhile, comparing with the case using only one key generation basis combination, the adoption of two key generation basis combinations can double the practical key generation rate, thereby slightly increasing the secure photon transmission distance. For example, under the practical key generation rate of 1 bit/s, the maximal photon transmission distance of the SPS DI QSS protocol with only one key generation basis combination is $d=$76.53 km, while with two key generation basis combinations, the secure photon transmission distance of the SPS DI QSS protocol reaches 81.54 km.

\section{SPS DI QSS using active improvement strategies}\label{Section4}
In the above simulation, we assume the local efficiency $\eta_l=100\%$. However, in practical experimental process, due to the influence of noise in storage and measurement, the photons will suffer from local loss, leading to $\eta_l<100\%$. The local loss will seriously weaken the nonlocal correlation among the measurement results, thus reducing the practical key generation rate and secure photon transmission distance. Therefore, DI-type protocols often require high-performance experimental equipment, which would increase the experimental difficulty. Based on Eq. (\ref{EC}), the threshold of $\eta_{l}$ for positive $E_c$ is about 95.58\%. Refs.~\cite{DIQSS2,DIQSS3} once adopted the postselection strategy and the advanced postselection strategy, which combines postselection and noise preprocessing strategies to further reduce the global efficiency $\eta$ threshold ($\eta=\eta_{t}\eta_{l}$). In this SPS DI QSS protocol, we also adopt the postselection strategy and the advanced postselection strategy to improve its performance in practical experimental conditions.

First, with the postselection strategy, we randomly label the detector no-click event $\perp$ as $+1$ or $-1$. Compared with Eq.~\eqref{QBERQ2}, the QBER $Q_2^p$ caused by photon loss after using the postselection strategy is
\begin{eqnarray}\label{pQBERQ2}
Q_2^p=\frac{1}{2}(1-\eta^3_l)
\end{eqnarray}
As a result, the postselection strategy reduces the total QBER $\delta_p$ with the form of
\begin{eqnarray}\label{ptotalQBER2}
\delta_p=\frac{1-F}{2}\eta_l^3+\frac{1-\eta^3_l}{2}=\frac{1}{2}-\frac{1}{2}\eta^3_lF.
\end{eqnarray}

In this way, the key uncorrelation of Bob and Charlie are changed to $H(A_1|B_1,C_1)_p=H(A_2|B_1,C_2)_p=h(\delta_p)$. The total key secrecy rate $H(A|E)_p$ for Eve has the same lower bound as Eq.~\eqref{keysecrecyrate}.

After substituting Eqs.~\eqref{noiseS} and \eqref{ptotalQBER2}, we obtain the lower bound of the secure key rate $R^p_\infty$ of SPS DI QSS with postselection strategy as
\begin{eqnarray}\label{prinfty}
R^p_\infty &=&\left(p^2+\bar{p}^2\right)\Big[H(A|E)_p-\big(\lambda H(A_1|B_1,C_1)_p\nonumber\\
&+&\bar{\lambda}H(A_2|B_1,C_2)_p\big)\Big]\nonumber\\
&\ge& \left(p^2+\bar{p}^2\right)\left[g\left(\tilde{E}_\lambda\left(2\sqrt{2}F\eta_l^3\right)\right)- h(\frac{1}{2}-\frac{1}{2}\eta^3_lF)\right].\nonumber\\
\end{eqnarray}

\begin{figure}
\includegraphics[scale=0.34]{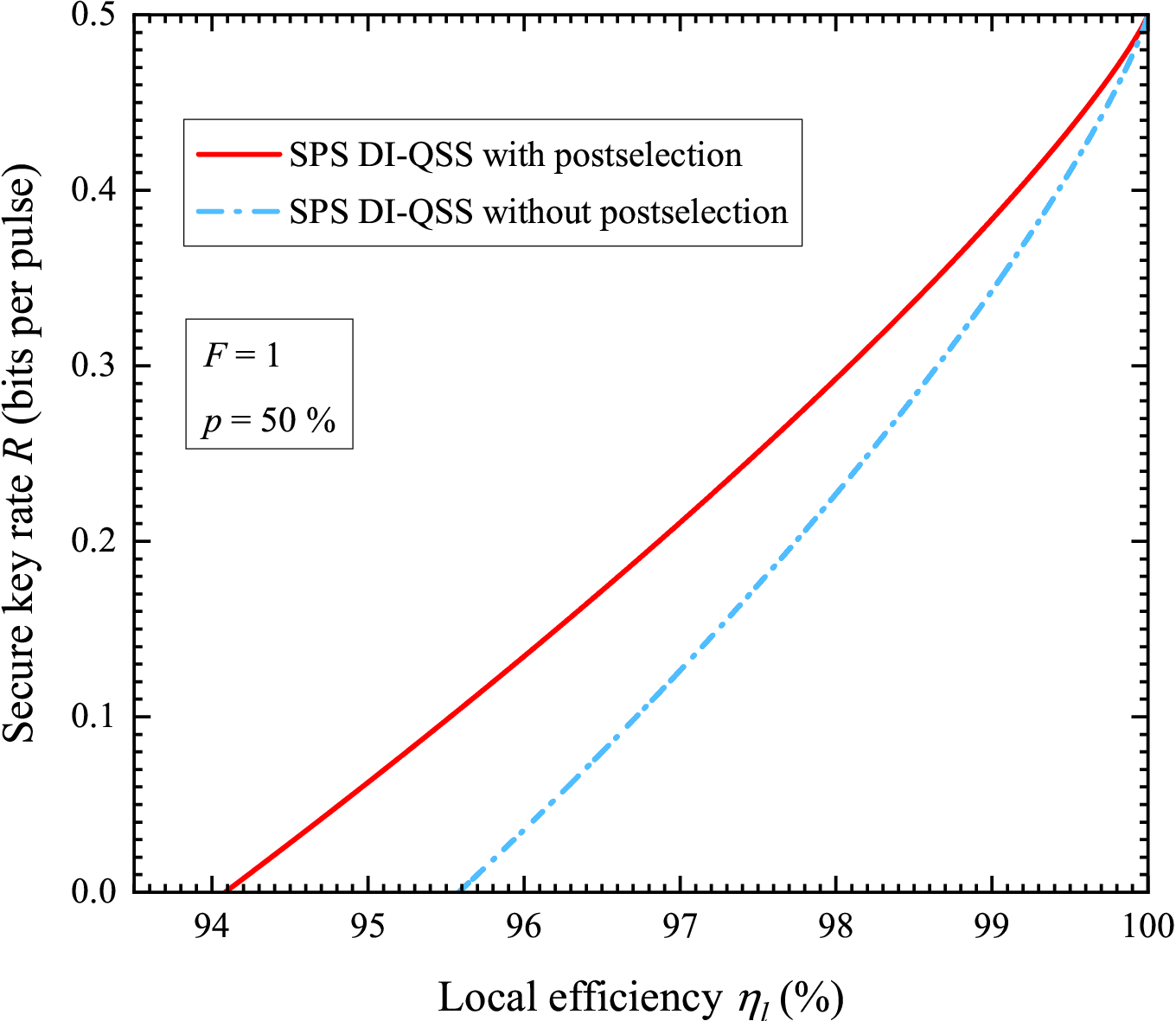}
\caption{(Color online) The secure key rate of SPS DI QSS altered with the local efficiency $\eta_{l}$ with $F=1$ and $p=50\%$.}
\label{fig4}
\end{figure}

In Fig.~\ref{fig4}, we compare the secure key rate $R^p_\infty$ of SPS DI QSS protocol with and without the postselection strategy altered with the local efficiency $\eta_{l}$ with $F=1$ and $p=50\%$. It can be seen that with the postselection strategy, the threshold of $\eta_l$ can be reduced from 95.58\% to 94.08\%.

To further deal with the qubit error rate caused by the postselection operation, we can adopt the noise preprocessing strategy. The noise preprocessing strategy is to add some artificial noise to the initial measurement data. When three users select the key generation basis combination $\{A_1B_1C_1\}$ and $\{A_2B_1C_2\}$, Alice flips her measurement results ($+1\leftrightarrow-1$) with probability $q$. In the error correction step, Alice announces the flip probability $q$. The users apply a hash function to the original key to obtain the final key. By performing the noise preprocessing strategy, the total QBER $\delta_q$ consists of two scenarios. Firstly, the initial measurement results do not suffer from the bit-flip error, but Alice flips her result with the probability of $q$. Secondly, the initial results suffer from the bit-flip error, and Alice does not flip her result with the probability of $\bar{q}=1-q$. Therefore, the total QBER $\delta_{qp}$ is given by
\begin{eqnarray}\label{qptotalQBER2}
\delta_{qp}=q+(1-2q)\delta_p=q+(1-2q)(\frac{1}{2}-\frac{1}{2}\eta^3_lF).
\end{eqnarray}
In this way, the key uncorrelation of Bob and Charlie are changed to $H(A_1|B_1,C_1)_{qp}=H(A_2|B_1,C_2)_{qp}=h(\delta_{qp})$.

According to the derivation of Ref.~\cite{DIQSS2,DIQSS3}, after performing the noise preprocessing operation, the lower bound of the total key secrecy rate for Eve can be calculated as
\begin{eqnarray}\label{qpkeysecrecyrate}
H\left(A|E\right)_q\ge g\left(\tilde{E}_\lambda(S),q\right),
\end{eqnarray}
where $\tilde{E}_\lambda(S)\equiv \sqrt{\tilde{E}_\lambda(S)^2}$, and $\tilde{E}_\lambda(S)^2$ is the maximum value obtained by numerically or analytically solving the optimization problem \eqref{maxproblem}. we define the function $g(x,q)$ as
\begin{eqnarray}\label{gxq}
g(x,q)&=&1-h\left(\frac{\sqrt{x^2/4-1}}{2}+\frac{1}{2}\right)\nonumber\\
&+&h\left[\frac{\sqrt{\left(1-2q\right)^2+4q\left(1-q\right)\left(x^2/4-1\right)}}{2}+\frac{1}{2}\right].\nonumber\\
\end{eqnarray}

According to Eq.~\eqref{rinfty2}, the lower bound of secure key rate $R^{qp}_\infty$ of SPS DI QSS with advanced postselection strategy as
\begin{eqnarray}\label{qprinfty}
R^{qp}_\infty &=&\left(p^2+\bar{p}^2\right)\Big[H(A|E)_{qp}-\big(\lambda H(A_1|B_1,C_1)_{qp}\nonumber\\
&+&\bar{\lambda}H(A_2|B_1,C_2)_{qp}\big)\Big]\nonumber\\
&\ge& \left(p^2+\bar{p}^2\right)\left[g\left(\tilde{E}_\lambda\left(S\right),q\right)- h(\delta_{qp})\right].
\end{eqnarray}

Substituting Eq.~\eqref{noiseS} and Eq.~\eqref{qptotalQBER2} into Eq.~\eqref{qprinfty}, the practical key generation rate $E_c^{qp}$ of SPS DI QSS with the advanced postselection strategy is given by
\begin{eqnarray}\label{qpEC}
E_c^{qp}&=&(1-P_c)R_{rep}P_dR^{qp}_\infty\nonumber\\
&\geq& \frac{\eta_t^3R_{rep}}{64}(p^2+\bar{p}^2)\Bigg[g\left(\tilde{E} _\lambda\left(2\sqrt{2}F\eta_l^3\right),q\right)\nonumber\\
&-&h\left(q+\left(1-2q\right)\left(\frac{1}{2}-\frac{1}{2}\eta^3_lF\right)\right)\Bigg].
\end{eqnarray}

\begin{figure}
\includegraphics[scale=0.34]{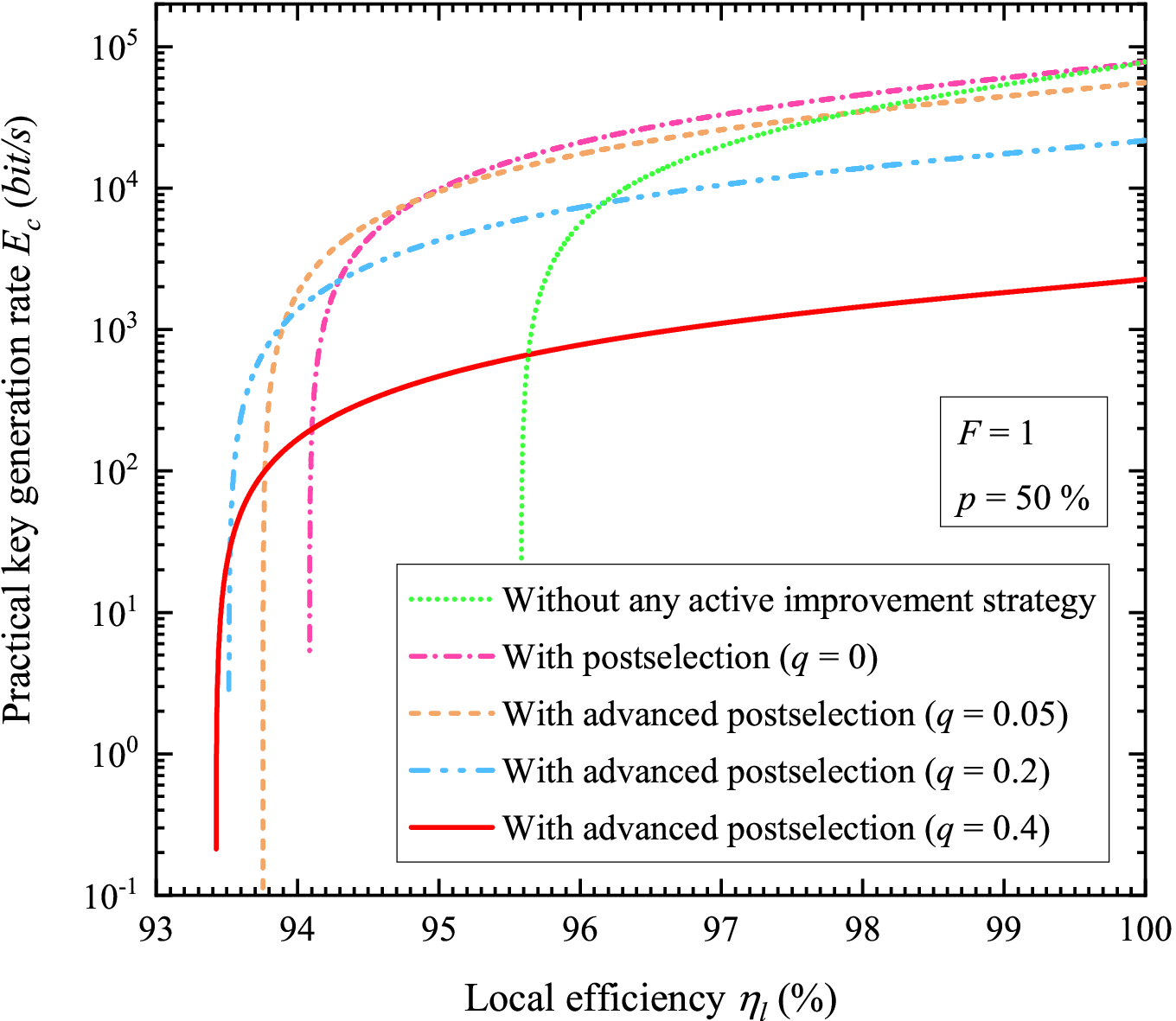}
\caption{(Color online) The practical key generation rate of SPS DI QSS without any active improvement strategy, with the postselection strategy ($q=0$) and the advanced postselection strategy ($q=$0.05, 0.2, 0.4) as a function of the local efficiency $\eta_l$, respectively. Here, we assume $F=1$ and $p=50\%$.}
\label{fig5}
\end{figure}

Fig.~\ref{fig5} shows the practical key generation rate of SPS DI QSS protocol without any active improvement strategy, with the postselection strategy and the advanced postselection strategy versus the local efficiency $\eta_l$, respectively. Here, we control $F=1$ and $p=50\%$, and set $q=$0, 0.05, 0.2, 0.4, where $q=0$ corresponds to the postselection strategy. It can be found that the protocol with the postselection strategy $q=0$ has the highest practical key generation rate. It seems that under the relatively high local efficiency $\eta_{l}$, the adoption of postselection strategy has a greater advantage in promoting the key generation. The practical key generation rate would reduce with the growth of $q$, but the growth of $q$ would decrease the local efficiency threshold. The reason is that although adding artificial noise to Alice's measurement results would increase the uncorrelation among three user's key bits (entropy $H(A_1|B_1,C_1)$ and $H(A_2|B_1,C_2)$), it can effectively improve the total key secrecy rate $H(A|E)$ to Eve. As a result, the net effect on the noise robustness is positive. When $q\rightarrow0.5$, the local efficiency threshold can be further decreased from 94.08\% to 93.41\%. As a result, the adoption of the advanced postselection strategy can further reduce the requirements for the performance of fiber, quantum memory, and photon detector, thus reducing the difficulty of the SPS DI QSS's experimental implementation.

\begin{figure}
\includegraphics[scale=0.34]{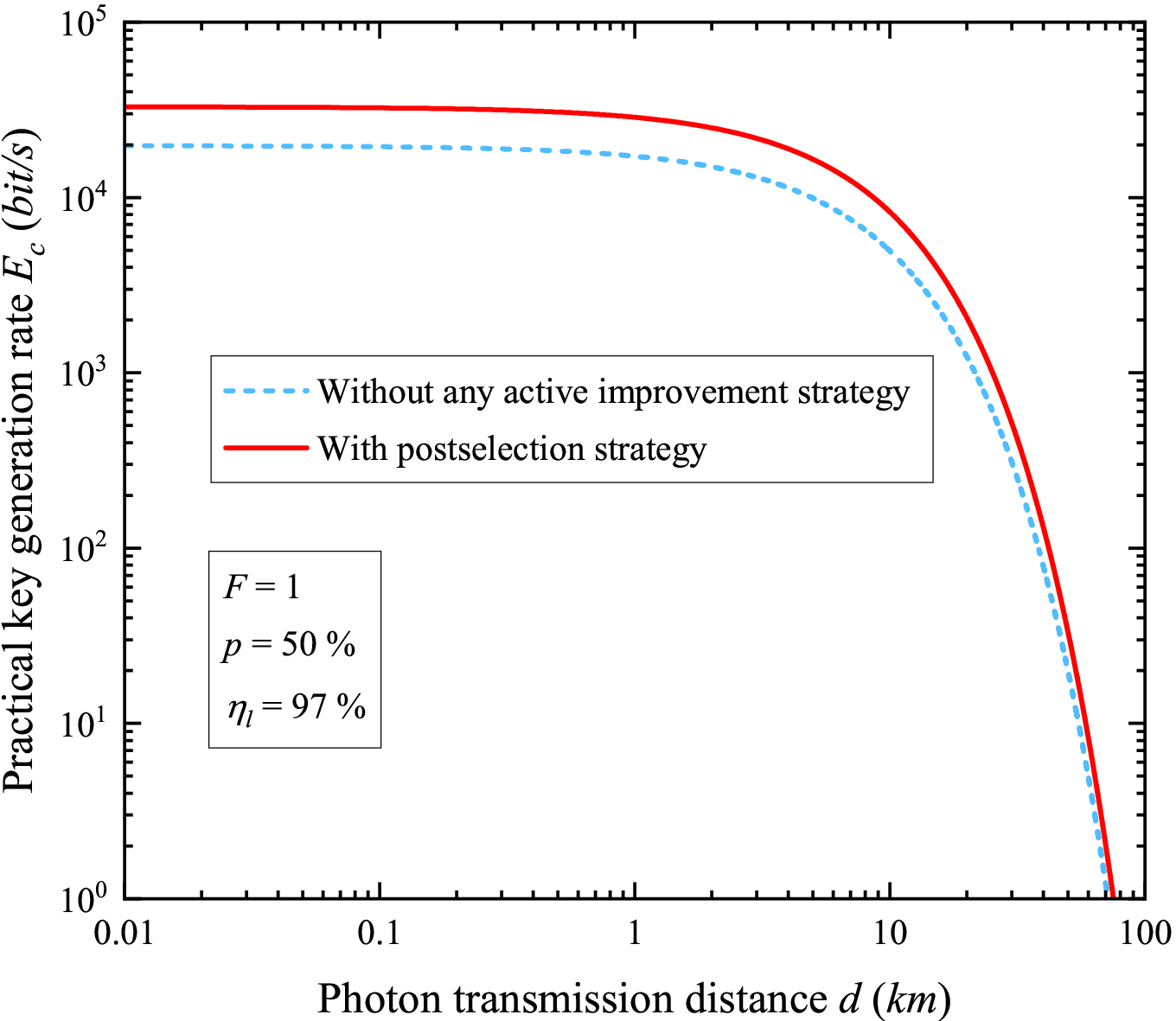}
\caption{(Color online) The practical key generation rate of SPS DI QSS with and without the postselection strategy as a function of the photon transmission distance $d$ under the local efficiency $\eta_l=97\%$. Here, we assume $F=1$ and $p=50\%$.}
\label{fig6}
\end{figure}

In Fig.~\ref{fig6}, we provide the practical key generation rate of SPS DI QSS without any active improvement strategy and with the postselection strategy as a function of the photon transmission distance $d$ under the local efficiency $\eta_l=97\%$. It can be seen that the adoption of the postselection strategy increases SPS DI QSS's practical key generation rate to 1.67 times of the original value. Meanwhile, under the practical key generation rate of 1 bit/s, the photon transmission distance of the protocol without the active improvement strategy is about 71.58 km, while that of the protocol with postselection strategy is increased to about 75.29 km.

\section{Discussion and conclusion}\label{Section5}
In our work, we propose the three-partite SPS DI QSS protocol and estimate its performance in the practical communication environment. The adoption of the high-efficient SPSs and the heralded structure can largely increase the practical key generation rate by six orders of magnitude and increase the secure photon transmission distance. Moreover, we adopt the active improvement strategies to further decrease the local efficiency threshold of SPS DI QSS, which can promote its experimental demonstration. Naturally, $m$ ($m>3$) users can also construct the long-distance entanglement channels with the help of single photon sources and the $m$-photon GSM. As a result, our three-partite SPS DI QSS protocol can be naturally extended to the arbitrary $m$-partite SPS DI QSS protocol \cite{DIQSS2,DIQSS3}, which has great potential to realize high-efficiency SPS DI quantum communication network.

In our paper, the key elements for constructing the long-distance multi-partite entanglement channels are the GSM module and the QM module. These modules can be realized with linear optical devices. Under linear optical conditions, the GSM module only correctly identifies two of the eight three-photon polarization GHZ states with the success probability of 1/4 \cite{GSM}. To improve the success probability of the GSM module, we can also adopt the GSM based on quantum non-destructive measurement \cite{GSM1} and hyperentanglement-assisted GSM \cite{GSM2} to achieve the complete discrimination of the eight GHZ states. These two GSM approaches are expected to increase the success probability of constructing entangled channels by four times. On the other hand, unlike general QMs based on the solid-states \cite{QM2,QM3,QM4,QM5}, we use the all-optical storage loop QM to implement the store and readout of photons. The all-optical storage loop QM was experimentally demonstrated with the storage efficiency of 91\% and a life time of 131 ns for photons with a central wavelength of 1550 nm and a bandwidth of 0.52 THz, corresponding to around 11 round-trips \cite{QM1}. The lifetime of 131 ns is sufficient for the subsequent GSM. Moreover, the storage loop QM can operate at any given wavelength in theory, with only minor adaptations. Meanwhile, with the advanced postselection strategy, the local efficiency threshold of the SPS DI QSS protocol is reduced to 93.41\% ($q\rightarrow0.5$). The requirement of the local efficiency $\eta_l$ depends on high-efficient single photon detectors. In 2020, the superconducting nanowire single photon detector reached the detection efficiency of 98\% in the 1550 nm band \cite{SNSPD}, which satisfies the detection efficiency threshold of SPS DI QSS.

In conclusion, we propose the high-efficient long-distance SPS DI QSS protocol in this paper. All the users prepare single photons with orthogonal polarization from SPSs. With the help of the path emerging and the VBS, they each send one photon for the multi-photon GSM and store the other photon in the QM module. When the GSM is successful and all the QMs are loaded by photons, all users' stored photons can deterministically construct the long-distance multi-partite entanglement. Then, each user extracts the stored photon and randomly chooses measurement basis to measure it. Our protocol chooses two measurement basis combinations to generate key and guarantee the security based on the violation of Svetlichny inequality. The players can finally read out the dealer's keys by cooperation. We develop a simulation method to estimate its practical key generation rate. Comparing with the previous SPDC DI QSS protocols \cite{DIQSS2,DIQSS3}, our SPS DI QSS protocol has some attractive features. First, the SPDC source generates entangled states with a quite low probability, but the practical SPS is nearly on-demand, which can largely increase the success probability of constructing the multi-partite entanglement channels. Second, the heralded function of the GSM and QM can automatically eliminate the negative influence from photon transmission loss on the Svetlichny polynomial, and thus largely increase the secure photon transmission distance. Third, by adopting the active improvement strategy, we can further increase the practical key generation rate of SPS DI QSS and reduce its local efficiency threshold (reduce the experimental difficulty). The SPS DI QSS protocol can largely increase the practical key generation rate by six orders of magnitude. In theory, it has an infinite secure photon transmission distance. Under the practical key generation rate of $10^{-4}$ bit/s, the secure photon transmission distance reaches 148.21 km (about 183 times of 0.81 km in Ref. \cite{DIQSS2} and 151 times of 0.98 km in Ref. \cite{DIQSS3}). Combined with the advanced postselection strategy, its local efficiency threshold is reduced to about 93.41\%.  Our three-partite SPS DI QSS protocol can be extended to the general multi-partite SPS DI QSS protocol and is feasible with the current experimental technologies. Our protocol makes it possible to realize high-efficient long distance DI network in the near future.

\section*{Acknowledgement}
This work was supported by the National Natural Science Foundation of China under Grants No. 12175106 and No. 92365110, the Natural Science Foundation of Jiangsu Province of China under Grants No. BK20240611 and No. BK20240612, the Postgraduate Research \& Practice Innovation Program of Jiangsu Province under Grant No.KYCX25-1245.


\begin{thebibliography}{26}
	
\bibitem{QSS1} M. Hillery, V. Bu\v{z}ek and A. Berthiaume, Quantum secret sharing, \href{https://doi.org/10.1103/PhysRevA.59.1829}{Phys. Rev. A \textbf{59}, 1829 (1999)}.
    
\bibitem{QSS2} A. Karlsson, M. Koashi and N. Imoto, Quantum entanglement for secret sharing and secret splitting, \href{https://doi.org/10.1103/PhysRevA.59.162}{Phys. Rev. A \textbf{59}, 162 (1999)}.
    
\bibitem{QSS3} R. Cleve, D. Gottesman and H.-K. Lo, How to share a quantum secret, \href{https://doi.org/10.1103/PhysRevLett.83.648}{Phys. Rev. Lett. \textbf{83}, 648 (1999)}.
    
\bibitem{QSS4} L. Xiao, G. L. Long, F. G. Deng and J. W. Pan, Efficient multiparty quantum-secret-sharing schemes, \href{https://doi.org/10.1103/PhysRevA.69.052307}{Phys. Rev. A \textbf{69}, 052307 (2004)}.
    
\bibitem{QSS5} Z. J. Zhang, Y. Li and Z. X. Man, Multiparty quantum secret sharing, \href{https://doi.org/10.1103/PhysRevA.71.044301}{Phys. Rev. A \textbf{71}, 044301 (2005)}.

\bibitem{QSS6} Z. J. Zhang and Z. X. Man, Multiparty quantum secret sharing of classical messages based on entanglement swapping, \href{https://doi.org/10.1103/PhysRevA.72.022303}{Phys. Rev. A \textbf{72}, 022303 (2005)}.

\bibitem{QSS7} D. Markham and B. C. Sanders, Graph states for quantum secret sharing, \href{https://doi.org/10.1103/PhysRevA.78.042309}{Phys. Rev. A \textbf{78}, 042309 (2008)}.

\bibitem{QSS8} A. Tavakoli, I. Herbauts, M. \.{Z}ukowski and M. Bourennane, Secret sharing with a single d-level quantum system, \href{https://doi.org/10.1103/PhysRevA.92.030302}{Phys. Rev. A \textbf{92}, 030302 (2015)}.

\bibitem{QSS9} W. P. Grice and B. Qi, Quantum secret sharing using weak coherent states, \href{https://doi.org/10.1103/PhysRevA.100.022339}{Phys. Rev. A \textbf{100}, 022339 (2019)}.

\bibitem{QSS10} B. P. Williams, J. M. Lukens, N. A. Peters, B. Qi and W. P. Grice, Quantum secret sharing with polarization-entangled photon pairs, \href{https://doi.org/10.1103/PhysRevA.99.062311}{Phys. Rev. A \textbf{99}, 062311 (2019)}.

\bibitem{QSS11} H. Wang, D. Liao, D. Guo, J. Xin and J. Kong, Continuous-variable (3, 3)-threshold quantum secret sharing based on one-sided device-independent security, \href{https://doi.org/10.1016/j.physleta.2023.128650}{Phys. Lett. A \textbf{462}, 128650 (2023)}.

\bibitem{MDIQSS1} Y. Fu, H. L. Yin, T. Y. Chen and Z. B. Chen, Long-distance measurement-device-independent multiparty quantum communication, \href{https://doi.org/10.1103/PhysRevLett.114.090501}{Phys. Rev. Lett. \textbf{114}, 090501 (2015)}.

\bibitem{MDIQSS2} X. X. Ju, W. Zhong, Y. B. Sheng and L. Zhou, Measurement-device-independent quantum secret sharing with hyper-encoding, \href{https://doi.org/10.1088/1674-1056/ac70bb}{Chin. Phys. B \textbf{31}, 100302 (2022)}.

\bibitem{DMDIQSS1} Z. K. Gao, T. Li and Z. H. Li, Deterministic measurement-device-independent quantum secret sharing, \href{https://doi.org/10.1007/s11433-020-1603-7}{Sci. China: Phys. Mech. Astron. \textbf{63}, 120311 (2020)}.

\bibitem{DPS1} J. Gu, X. Y. Cao, H. L. Yin and Z. B. Chen, Differential phase shift quantum secret sharing using a twin field, \href{https://doi.org/10.1364/OE.417856}{Opt. Express \textbf{29}, 9165 (2021)}.

\bibitem{RR1} J. Gu, Y. M. Xie, W. B. Liu, Y. Fu, H. L. Yin and Z. B. Chen, Secure quantum secret sharing without signal disturbance monitoring, \href{https://doi.org/10.1364/OE.440365}{Opt. Express \textbf{29}, 32244 (2021)}.

\bibitem{QSS12} Y. Ouyang, K. Goswami, J. Romero, B. C. Sanders, M. H. Hsieh and M. Tomamichel, Approximate reconstructability of quantum states and noisy quantum secret sharing schemes, \href{https://doi.org/10.1103/PhysRevA.108.012425}{Phys. Rev. A \textbf{108}, 012425 (2023)}.

\bibitem{QSSe1} Y. A. Chen, A. N. Zhang, Z. Zhao, X. Q. Zhou, C. Y. Lu, C. Z. Peng, T. Yang and J. W. Pan, Experimental quantum secret sharing and third-man quantum cryptography, \href{https://doi.org/10.1103/PhysRevLett.95.200502}{Phys. Rev. Lett. \textbf{95}, 200502 (2005)}.

\bibitem{QSSe2} S. Gaertner, C. Kurtsiefer, M. Bourennane and H. Weinfurter, Experimental demonstration of four-party quantum secret sharing, \href{https://doi.org/10.1103/PhysRevLett.98.020503}{Phys. Rev. Lett. \textbf{98}, 020503 (2007)}.

\bibitem{QSSe3} H. Lu, Z. Zhang, L. K. Chen, Z. D. Li, C. Liu, L. Li, N. L. Liu, X. F. Ma, Y. A. Chen and J. W. Pan, Secret sharing of a quantum state, \href{https://doi.org/10.1103/PhysRevLett.117.030501}{Phys. Rev. Lett. \textbf{117}, 030501 (2016)}.

\bibitem{QSSe4} Y. Y. Zhou, J. Yu, Z. H. Yan, X. J. Jia, J. Zhang, C. D. Xie and K. C. Peng, Quantum secret sharing among four players using multipartite bound entanglement of an optical field, \href{https://doi.org/10.1103/PhysRevLett.121.150502}{Phys. Rev. Lett. \textbf{121}, 150502 (2018)}.

\bibitem{QSSe5} C. Schmid, P. Trojek, M. Bourennane, C. Kurtsiefer, M. \.{Z}ukowski and H. Weinfurter, Experimental single qubit quantum secret sharing, \href{https://doi.org/10.1103/PhysRevLett.95.230505}{Phys. Rev. Lett. \textbf{95}, 230505 (2005)}.

\bibitem{QSSe6} B. A. Bell, D. Markham, M. D. A. Herrera, A. Marin, W. J. Wadsworth, J. G. Rarity and M. S. Tame, Experimental demonstration of graph-state quantum secret sharing, \href{https://doi.org/10.1038/ncomms6480}{Nat. Commun. \textbf{5}, 5480 (2014)}.

\bibitem{QSSe7} Y. Cai, J. Roslund, G. Ferrini, F. Arzani, X. Xu, C. Fabre and N. Treps, Multimode entanglement in reconfigurable graph states using optical frequency combs, \href{https://doi.org/10.1038/ncomms15645}{Nat. Commun. \textbf{8}, 15645 (2017)}.

\bibitem{QSSe8} A. Shen, X. Y. Cao, Y. Wang, Y. Fu, J. Gu, W. B. Liu, C. X. Weng, H. L. Yin and Z. B. Chen, Experimental quantum secret sharing based on phase encoding of coherent states, \href{https://doi.org/10.1007/s11433-023-2105-7}{Sci. China: Phys. Mech. Astron. \textbf{66}, 143 (2023)}.

\bibitem{Bell} J. S. Bell, On the Einstein Podolsky Rosen paradox, \href{https://doi.org/10.1103/PhysicsPhysiqueFizika.1.195}{Phys. Phys. Fizika \textbf{1}, 195 (1964)}.

\bibitem{CHSH} J. F. Clauser, M. A. Horne, A. Shimony and R. A. Holt, Proposed experiment to test local hidden-variable theories, \href{https://doi.org/10.1103/PhysRevLett.23.880}{Phys. Rev. Lett. \textbf{23}, 880 (1969)}.

\bibitem{DIQRG1} Y. Liu, Q. Zhao, M. H. Li, \emph{et al.,} Device-independent quantum random-number generation, \href{https://doi.org/10.1038/s41586-018-0559-3}{Nature \textbf{562}, 548-551 (2018)}.

\bibitem{DIQRG2} M. H. Li, X. J. Zhang, W. Z. Liu, \emph{et al.,} Experimental realization of device-independent quantum randomness expansion, \href{https://doi.org/10.1103/PhysRevLett.126.050503}{Phys. Rev. Lett. \textbf{126}, 050503 (2021)}.

\bibitem{DIQKD} A. Ac\'{\i}n, N. Gisin, and L. Masanes, From Bell's theorem to secure quantum key distribution, \href{https://doi.org/10.1103/PhysRevLett.97.120405}{Phys. Rev. Lett. \textbf{97}, 120405 (2006)}.

\bibitem{DIQKD1} A. Ac\'{\i}n, N. Brunner, N. Gisin, S. Massar, S. Pironio and V. Scarani, Device-independent security of quantum cryptography against collective attacks, \href{https://doi.org/10.1103/PhysRevLett.98.230501}{Phys. Rev. Lett. \textbf{98}, 230501 (2007)}.

\bibitem{DIQKD2} S. Pironio, A. Ac\'{\i}n, N. Brunner, N. Gisin, S. Massar and V. Scarani, Device-independent quantum key distribution secure against collective attacks, \href{https://doi.org/10.1088/1367-2630/11/4/045021}{New J. Phys. \textbf{11}, 045021 (2009)}.

\bibitem{DIQKD3} L. Masanes, S. Pironio and A. Ac\'{\i}n, Secure device-independent quantum key distribution with causally independent measurement devices, \href{https://doi.org/10.1038/ncomms1244}{Nat. Commun. \textbf{2}, 238 (2011)}.

\bibitem{DIQKD4} C. C. W. Lim, C. Portmann, M. Tomamichel, R. Renner and N. Gisin, Device-independent quantum key distribution with local Bell test, \href{https://doi.org/10.1103/PhysRevX.3.031006}{Phys. Rev. X \textbf{3}, 031006 (2013)}.

\bibitem{DIQKD5} U. Vazirani and T. Vidick, Fully device-independent quantum key distribution, \href{https://doi.org/10.1103/PhysRevLett.113.140501}{Phys. Rev. Lett. \textbf{113}, 140501 (2014)}.

\bibitem{DIQKD6} X. F. Ma and N. L\"{u}tkenhaus, Improved data post-processing in quantum key distribution and application to loss thresholds in device independent QKD, \href{https://doi.org/10.26421/qic12.3-4-2}{Quantum Info. Comput. \textbf{12}, 203 (2012)}.
    
\bibitem{DIQKD7} L. P. Thinh, G. de la Torre, J.-D. Bancal, S. Pironio and V. Scarani, Randomness in post-selected events, \href{https://doi.org/10.1088/1367-2630/18/3/035007}{New J. Phys. \textbf{18}, 035007 (2016)}.

\bibitem{DIQKD8} M. Ho, P. Sekatski, E. Y. Z. Tan, R. Renner, J.-D. Bancal and N. Sangouard, Noisy preprocessing facilitates a photonic realization of device-independent quantum key distribution, \href{https://doi.org/10.1103/PhysRevLett.124.230502}{Phys. Rev. Lett. \textbf{124}, 230502 (2020)}.

\bibitem{DIQKD9} E. Woodhead, A. Ac\'{\i}n and S. Pironio, Device-independent quantum key distribution with asymmetric CHSH inequalities, \href{https://doi.org/10.22331/q-2021-04-26-443}{Quantum \textbf{5}, 443 (2021)}.

\bibitem{DIQKD10} P. Sekatski, J.-D. Bancal, X. Valcarce, E. Y. Z. Tan, R. Renner and N. Sangouard, Device-independent quantum key distribution from generalized CHSH inequalities, \href{https://doi.org/10.22331/q-2021-04-26-444}{Quantum \textbf{5}, 444 (2021)}.

\bibitem{DIQKD11} M. Masini, S. Pironio and E. Woodhead, Simple and practical DIQKD security analysis via BB84-type uncertainty relations and Pauli correlation constraints, \href{https://doi.org/10.22331/q-2022-10-20-843}{Quantum \textbf{6}, 843 (2022)}.

\bibitem{DIQKD12} F. H. Xu, Y. Z. Zhang, Q. Zhang and J. W. Pan, Device-independent quantum key distribution with random postselection, \href{https://doi.org/10.1103/PhysRevLett.128.110506}{Phys. Rev. Lett. \textbf{128}, 110506 (2022)}.

\bibitem{DIQKD13} Q. Zeng, H. Wang, H. Yuan, Y. Fan, L. Zhou, Y. Gao, H. Ma and Z. Yuan, Controlled entanglement source for quantum cryptography, \href{https://doi.org/10.1103/PhysRevApplied.19.054048}{Phys. Rev. Appl. \textbf{19}, 054048 (2023)}.

\bibitem{DIQKD14} R. Schwonnek, K. T. Goh, I. W. Primaatmaja, E. Y.-Z. Tan, R. Wolf, V. Scarani and C. C.-W. Lim, Device-independent quantum key distribution with random key basis, \href{https://doi.org/10.1038/s41467-021-23147-3}{Nat. Commun. \textbf{12}, 2880 (2021)}.

\bibitem{DIQKD15} E. Y. Z. Tan, P. Sekatski, J. D. Bancal, R. Schwonnek, R. Renner, N. Sangouard and C. C. W. Lim, Improved DIQKD protocols with finite-size analysis, \href{https://doi.org/10.22331/q-2022-12-22-880}{Quantum \textbf{6}, 880 (2022)}.

\bibitem{DIQKD16} E. Y. Z. Tan, C. C.-W. Lim and R. Renner, Advantage distillation for device-independent quantum key distribution, \href{https://doi.org/10.1103/PhysRevLett.124.020502}{Phys. Rev. Lett. \textbf{124}, 020502 (2020)}.

\bibitem{DIQKD17} Y. Z. Zhen, Y. Mao, Y. Z. Zhang, F. Xu and B. C. Sanders, Device-independent quantum key distribution based on the Mermin-Peres magic square game, \href{https://doi.org/10.1103/PhysRevLett.131.080801}{Phys. Rev. Lett. \textbf{131}, 080801 (2023)}.
    

\bibitem{Heralded1} J. Kołody\'{n}ski, A. Máttar, P. Skrzypczyk, E. Woodhead, D. Cavalcanti, K. Banaszek and A. Ac\'{\i}n, Device-independent quantum key distribution with single-photon sources, \href{https://doi.org/10.22331/q-2020-04-30-260}{Quantum \textbf{4}, 260 (2020)}.
    
\bibitem{Heralded3} E. M. González-Ruiz, J. Rivera-Dean, M. F. Cenni, A. S. Sørensen, A. Acín and E. Oudot, Device-independent quantum key distribution with realistic single-photon source implementations, \href{https://doi.org/10.1364/OE.497935}{Opt. Express \textbf{32}, 13181 (2024)}.
    
    
\bibitem{DIQSS1} S. Roy and S. Mukhopadhyay, Device-independent quantum secret sharing in arbitrary even dimensions, \href{https://doi.org/10.1103/PhysRevA.100.012319}{Phys. Rev. A \textbf{100}, 012319 (2019)}.

\bibitem{DIQSS2} Q. Zhang, W. Zhong, M. M. Du, S. T. Shen, X. Y. Li, A. L. Zhang, L. Zhou and Y. B. Sheng, Device-independent quantum secret sharing with noise preprocessing and postselection, \href{https://doi.org/10.1103/PhysRevA.110.042403}{Phys. Rev. A \textbf{110}, 042403 (2024)}.

\bibitem{DIQSS3} Q. Zhang, J. W. Ying, Z. J. Wang, W. Zhong, M. M. Du, S. T. Shen, X. Y. Li, A. L. Zhang, S. P. Gu, X. F. Wang, \emph{et al.}, Device-independent quantum secret sharing with random key basis, \href{https://doi.org/10.1103/PhysRevA.111.012603}{Phys. Rev. A \textbf{111}, 012603(2025)}.

\bibitem{DIQSDC1} L. Zhou, Y. B. Sheng and G. L. Long, Device-independent quantum secure direct communication against collective attacks, \href{https://doi.org/10.1016/j.scib.2019.10.025}{Sci. Bull. \textbf{65}, 12 (2020)}.
    
\bibitem{DIQSDC2} L. Zhou and Y. B. Sheng, One-step device-independent quantum secure direct communication, \href{https://doi.org/10.1007/s11433-021-1863-9}{Sci. China: Phys. Mech. Astron. \textbf{65}, 250311 (2022)}.
    
\bibitem{Heralded2} L. Zhou, B. W. Xu, W. Zhong and Y. B. Sheng, Device-independent quantum secure direct communication with single-photon sources, \href{https://doi.org/10.1103/PhysRevApplied.19.014036}{Phys. Rev. Appl. \textbf{19}, 014036 (2023)}.

\bibitem{DIQKDe1} W. Z. Liu, Y. Z. Zhang, Y. Z. Zhen, M. H. Li, Y. Liu, J. Fan, F. Xu, Q. Zhang and J. W. Pan, Toward a photonic demonstration of device-independent quantum key distribution, \href{https://doi.org/10.1103/PhysRevLett.129.050502}{Phys. Rev. Lett. \textbf{129}, 050502 (2022)}.

\bibitem{DIQKDe2} D. P. Nadlinger, P. Drmota, B. C. Nichol, G. Araneda, D. Main, R. Srinivas, D. M. Lucas, C. J. Ballance, K. Ivanov and E. Y.-Z. Tan, Experimental quantum key distribution certified by Bell's theorem, \href{https://doi.org/10.1038/s41586-022-04941-5}{Nature \textbf{607}, 682 (2022)}.
    
\bibitem{DIQKDe3} W. Zhang, T. van Leent, K. Redeker, R. Garthoff, R. Schwonnek, F. Fertig, S. Eppelt, W. Rosenfeld, V. Scarani and C. C.-W. Lim, A device-independent quantum key distribution system for distant users, \href{https://doi.org/10.1038/s41586-022-04891-y}{Nature \textbf{607}, 687 (2022)}.

\bibitem{Svetlichny} G. Svetlichny, Distinguishing three-body from two-body nonseparability by a Bell-type inequality, \href{https://doi.org/10.1103/PhysRevD.35.3066}{Phys. Rev. D \textbf{35}, 3066 (1987)}.

\bibitem{GHZ1} D. R. Hamel, L. K. Shalm, H. H\"{u}bel, A. J. Miller, F. Marsili, V. B. Verma, R. P. Mirin, S. W. Nam, K. J. Resch and T. Jennewein, Direct generation of three-photon polarization entanglement, \href{https://doi.org/10.1038/nphoton.2014.218}{Nat. Photonics \textbf{8}, 801 (2014)}.

\bibitem{GHZ2} X. M. Hu, C. X. Huang, Y. B. Sheng, L. Zhou, B. H. Liu, Y. Guo, C. Zhang, W. B. Xing, Y. F. Huang, C. F. Li, and G. C. Guo, Long-distance entanglement purification for quantum communication, \href{https://doi.org/10.1103/PhysRevLett.126.010503}{Phys. Rev. Lett. \textbf{126}, 010503 (2021)}.
    

\bibitem{source} M. Müller, S. Bounouar, K. D. Jöns, M. Glässl and P. Michler, On-demand generation of indistinguishable polarization-entangled photon pairs, \href{https://doi.org/10.1038/nphoton.2013.377}{Nat. Photonics \textbf{8}, 224 (2014)}.

\bibitem{source1} J. Claudon, J. Bleuse, N. S. Malik, M. Bazin, P. Jaffrennou, N. Gregersen, C. Sauvan, P. Lalanne and J.-M. Gérard, A highly efficient single-photon source based on a quantum dot in a photonic nanowire, \href{https://doi.org/10.1038/nphoton.2009.287x}{Nat. Photonics \textbf{4}, 174 (2010)}.
    
\bibitem{source3} N. Somaschi, V. Giesz, L. De Santis, J. Loredo, M. P. Almeida, G. Hornecker, S. L. Portalupi, T. Grange, C. Anton and J. Demory, Near-optimal single-photon sources in the solid state, \href{https://doi.org/10.1038/nphoton.2016.23}{Nat. Photonics \textbf{10}, 340 (2016)}.

\bibitem{source4} X. Ding, Y. He, Z. C. Duan, N. Gregersen, M. C. Chen, S. Unsleber, S. Maier, C. Schneider, M. Kamp and S. Höfling, On-demand single photons with high extraction efficiency and near-unity indistinguishability from a resonantly driven quantum dot in a micropillar, \href{https://doi.org/10.1103/PhysRevLett.116.020401}{Phys. Rev. Lett. \textbf{116}, 020401 (2016)}.

\bibitem{source2} A. Barbiero, J. Huwer, J. Skiba-Szymanska, D. J. Ellis, R. M. Stevenson, T. Müller, G. Shooter, L. E. Goff, D. A. Ritchie and A. J. Shields, High-performance single-photon sources at telecom wavelength based on broadband hybrid circular Bragg gratings, \href{https://doi.org/10.1021/acsphotonics.2c00810}{ACS Photonics \textbf{9}, 3060 (2022)}.

\bibitem{Heralded6} M. Lasota, C. Radzewicz, K. Banaszek and R. Thew, Linear optics schemes for entanglement distribution with realistic single-photon sources, \href{https://doi.org/10.1103/PhysRevA.90.033836}{Phys. Rev. A \textbf{90}, 033836 (2014)}.

\bibitem{QM1} E. Meyer-Scott, N. Prasannan, I. Dhand, C. Eigner, V. Quiring, S. Barkhofen, B. Brecht, M. B. Plenio and C. Silberhorn, Scalable generation of multiphoton entangled states by active feed-forward and multiplexing, \href{https://doi.org/10.1103/PhysRevLett.129.150501}{Phys. Rev. Lett. \textbf{129}, 150501 (2022)}.

\bibitem{GSM} J. W. Pan and A. Zeilinger, Greenberger-horne-zeilinger-state analyzer, \href{https://doi.org/10.1103/PhysRevA.57.2208}{Phys. Rev. A 57, 2208 (1998)}.
    
\bibitem{nonlocal1} J. D. Bancal, N. Brunner, N. Gisin and Y. C. Liang, Detecting genuine multipartite quantum nonlocality: a simple approach and generalization to arbitrary dimensions, \href{https://doi.org/10.1103/PhysRevLett.106.020405}{Phys. Rev. Lett. \textbf{106}, 020405 (2011)}.
    
\bibitem{DWrate1} I. Devetak and A. Winter, Relating quantum privacy and quantum coherence: an operational approach, \href{https://doi.org/10.1103/PhysRevLett.93.080501}{Phys. Rev. Lett. \textbf{93}, 080501 (2004)}.

\bibitem{DWrate2} R. Augusiak and P. Horodecki, Multipartite secret key distillation and bound entanglement, \href{https://doi.org/10.1103/PhysRevA.80.042307}{Phys. Rev. A \textbf{80}, 042307 (2009)}.  
    
\bibitem{GSM1} J. Qian, X. L. Feng and S. Q. Gong, Universal Greenberger-Horne-Zeilinger-state analyzer based on two-photon polarization parity detection, \href{https://doi.org/10.1103/PhysRevA.72.052308}{Phys. Rev. A \textbf{72}, 052308 (2005)}.

\bibitem{GSM2} S. Y. Song, Y. Cao, Y. B. Sheng and G. L. Long, Complete Greenberger–Horne–Zeilinger state analyzer using hyperentanglement, \href{https://doi.org/10.1007/s11128-012-0375-x}{Quantum Inf. Process. \textbf{12}, 381 (2013)}.


\bibitem{QM2} D. S. Ding, Z. Y. Zhou, B. S. Shi and G. C. Guo, Single-photon-level quantum image memory based on cold atomic ensembles, \href{https://doi.org/10.1038/ncomms3527}{Nat. Commun. \textbf{4}, 2527 (2013)}.

\bibitem{QM3} E. Distante, P. Farrera, A. Padr\'{o}n-Brito, D. Paredes-Barato, G. Heinze and H. De Riedmatten, Storing single photons emitted by a quantum memory on a highly excited Rydberg state, \href{https://doi.org/10.1038/ncomms14072}{Nat. Commun. \textbf{8}, 14072 (2017)}.

\bibitem{QM4} P. Vernaz-Gris, K. Huang, M. Cao, A. S. Sheremet, and J. Laurat, Highly-efficient quantum memory for polarization qubits in a spatially-multiplexed cold atomic ensemble,  \href{https://doi.org/10.1038/s41467-017-02775-8}{Nat. Commun. \textbf{9}, 363 (2018)}.

\bibitem{QM5}  Y. Wang, J. Li, S. Zhang, K. Su, Y. Zhou, K. Liao, S. Du, H. Yan, and S. L. Zhu, Efficient quantum memory for single-photon polarization qubits, \href{https://doi.org/10.1038/s41566-019-0368-8}{Nat. Photon. \textbf{13}, 346 (2019)}.

\bibitem{SNSPD} D. V. Reddy, R. R. Nerem, S. W. Nam, R. P. Mirin and V. B. Verma, Superconducting nanowire single-photon detectors with 98\% system detection efficiency at 1550 nm, \href{https://doi.org/10.1364/OPTICA.400751}{Optica \textbf{7}, 1649 (2020)}.

\end{thebibliography}
\end{document}